\documentclass[aps,prb,superscriptaddress,twocolumn,letter,citeautoscript]{revtex4}
\usepackage[dvips]{graphicx}
\usepackage{times}

\begin{document}
\title{Highly tunable spin-dependent electron transport through carbon atomic chains
connecting two zigzag graphene nanoribbons}

\author{Yuehua Xu}
\author{Bao-Ji Wang}
\affiliation{Key Laboratory of Advanced Microstructured Materials, MOE,
Department of Physics, Tongji University, 1239 Siping Road, Shanghai 200092, P.
R. of China}
\author{San-Huang Ke}
\email[Corresponding author, E-mail: ]{shke@tongji.edu.cn}
\affiliation{Key Laboratory of Advanced Microstructured Materials, MOE,
Department of Physics, Tongji University, 1239 Siping Road, Shanghai 200092, P. R. of China}
\affiliation{Beijing Computational Science Research Center, 3 Heqing Road,
Beijing 100084, P. R. of China}

\begin{abstract}
Motivated by recent experiments of successfully carving out stable carbon atomic chains from graphene
we investigate a device structure of a carbon chain connecting two zigzag
graphene nanoribbons with highly tunable spin-dependent transport properties.
Our calculation based on the non-equilibrium Green's function approach combined
with the density functional theory shows that the transport behavior is
sensitive to the spin configuration of the leads and the bridge position in the
gap. A bridge in the middle gives an overall good coupling except for around
the Fermi energy where the leads with anti-parallel spins create a small
transport gap while the leads with parallel spins give a finite density of
states and induce an even-odd oscillation in conductance in terms of the number
of atoms in the carbon chain. On the other hand, a bridge at the edge shows
a transport behavior associated with the spin-polarized edge states, presenting
sharp pure $\alpha$-spin and $\beta$-spin peaks beside the Fermi energy in the
transmission function.  
This makes it possible to realize on-chip interconnects or spintronic devices by
tuning the spin state of the leads and the bridge position. 
\end{abstract}
\maketitle

\section{Introduction}

Carbon based nanostructures, especially, quasi-1D structures like carbon
nanotubes (CNTs) and, recently, graphene nanoribbons (GNRs), are playing more
and more important role in the development of nanoelectronics
\cite{avouris2007carbon}, possibly leading to an era of carbon-based
electronics. The practical application of CNTs is, however, limited by the
challenge in controlling experimentally their diameter and chirality which
control whether they are metallic or semiconducting.
Different from CNTs, the electronic properties of GNRs are determined by their
edge geometry and width and have shown promise for future generation of
transistor \cite{PhysRevB.54.17954,RevModPhys.81.109, 
geim2007rise, PhysRevLett.98.206805, PhysRevLett.97.216803, PhysRevB.73.195411,
PhysRevLett.100.206803,doi:10.1021/nl070133j, doi:10.1021/nl801774a}.
Zigzag-edged GNRs (ZGNRs) are particularly intriguing because of their
spin-polarized edge states which are localized around the two edges. The
resulting spin-dependent transport may make them promising candidates for
applications in spintronics. 
In narrow ZGNRs with $n$ zigzag carbon chains (denoted by $n$-ZGNRs, $n$
$\lesssim 32$) the anti-parallel spin configuration is the ground state, which
is slightly more favorable in energy than the parallel-spin one, while in wider
$n$-ZGNRs ($n$ $\gtrsim 32$) the two spin configurations are both possible to
exist due to the negligible interaction between the two edge states
\cite{son2006half}. Even in narrow ZGNRs, the small energy difference makes it
possible to change the ground state from anti-parallel spins to parallel spins
by appling a magnetic or electric field \cite{son2006half}. 

Recently, an interesting progress related to graphene is the successful
fabrication of free-standing linear carbon atomic chains carved out from a
graphene sheet by high-energy electron beam
\cite{meyer2008imaging,chuvilin2009graphene,jin2009deriving}, which is found
stable and is connected by $sp^2$ bonding. Unlike CNTs and GNRs, a carbon atomic
chain has no chirality and width. Therefore, it provides an ideal transport
channel for molecular devices. 
Experimentally, a carbon chain made in this way can be used as an on-chip device
with the advantage of the perfect $sp^2$ connection to the leads already set.
This is in striking contrast to the situation in conventional molecular
electronics using metal electrodes where a well-defined molecule-lead contact
with a good reproducebility is a big challenge
\cite{Basch051668,Venkataraman06458,Ke05074704}. For the miniatrization of the
whole device, the two graphene leads can be cut to form GNRs. 
Theoretically, the device structure of a carbon chain connecting two ZGNR leads
is particularly interesting since it combines the simple transport channel with
the rich electronic properties of ZGNRs via the perfect $sp^2$ contact. Its
transport properties can be artificially tailored to realize different
functionalities, as shown in this work. First, the tunable spin state of the
leads can be used to control the density of states at the Fermi energy, giving
either a large or zero equilibrium conductance (switch). Second, the atomic thin
bridge can be used to explore the locally spin-polarized edge states, implying
that the transport behavior will be sensitive to the position of the bridge.
This may be used to realize selectively the functionality of either on-chip
interconnects or spintronics. 


So far, the conductance of carbon chains connected to different types of
electrodes have recently been studied by several research groups
\cite{tongay2004ab, lang2000carbon, zhou2008first, brandbyge2002density,
larade2001conductance,wei2004spin, khoo2008negative, 
cheraghchi2008negative,furst2010atomic} but how it behaves is still an open
problem depending on the nature of the electrode used. Among these work, a
calculation \cite{furst2010atomic} of spin-denpendent transport was reported for
a junction with very narrow ZGNR leads connected via a spin-polarized 5-member
ring which is, however, unlikely to be carved out directly from a graphene sheet
and is probably unstable due to the unpaired $p_z$ electron without doping \cite{Ke07146802}.

Recently Shen {\it et al} reported a theoretical calculation for carbon
chain-ZGNR junctions with very narrow ZGNR leads but ignoring totally the spin freedom
\cite{shen2010electron}. An interesting finding from their 
calculation is that the equilibrium conductance shows an even-odd oscillation
with regard to the number of atoms in the carbon chain \cite{shen2010electron}:
A chain with an odd number of atoms will have a larger conductance than one with
an even number of atoms. This even-odd behavior is due to the very
sharp peak in the transmission function at the Fermi energy, whose origination
was not understood clearly and was ascribed to the edge states of the leads.
However, if one takes the spin freedom into account,
this even-odd behavior will totally disappear for the ground state since the
anti-parallel spins will create a band gap in the leads. 

Another recent work about transport properties of carbon chain-GNR junctions was
reported by Zanolli {\it et al}\cite{zanolli2010quantum}, in which the spin
freedom was taken into account. In their calculation wider GNR leads are considered
and a carbon chain is connected right in the middle of the gap. 
It was found that the spin freedom has a significant effect. Especially, an
odd-numbered carbon chain bridge is found to be spin-polarized by itself while
an even-numbered one is not. This calculation showed that an odd-numbered carbon
chain combined with parallel-spin leads can give a large spin-polarized
equilibrium conductance due to the peaks in the transmission function around the
Fermi energy. This result implys that these peaks are not associated with the edge states
because the bridge is in the middle and far away from the edges
though their origination was unclear.


Despite these previous studies, a full understanding about the transport
properties of carbon chain-ZGNR junctions is still lacking, especailly the
effect from the bridge position and its combination
with the spin configuration and width of the leads, as
well as the underlying physics leading to the transport behavior.
In this work, we investigate the spin-dependent electron transport of 
carbon chain-ZGNR junctions by performing first-principles calculations based on
the non-equilibrium Green's function (NEGF) approach combined with the density
functional theory (DFT) \cite{datta95, ke2004electron}. We study systematically
the effects from all the factors mentioned above. 
Our calculation shows that the bridge position combined with the
spin freedom of the leads dominates the transport behavior. The uderlying
physics leading to this transport behavior is fully analyzed by investigating the
space-resolved density of states (DOS). 
%
%
The present work indicates that different functionalities can be achieved by
changing the bridge position and the spin state of the leads: on-chip
interconnects can be realized with a bridge connected in the middle and a
parallel spin configuration of the leads, which give a large equilibrium
conductance, while a device for spintronics may be realized with a bridge
connected at the edge, which gives a very high spin-polarization ratio. 

\begin{figure}[b]
(a) \includegraphics[width=7.3cm,clip]{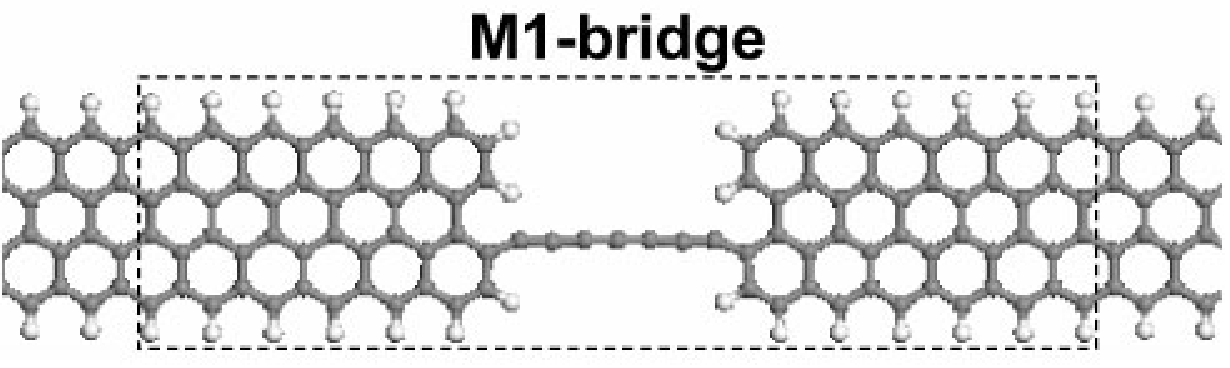} \\
(b) \includegraphics[width=7.3cm,clip]{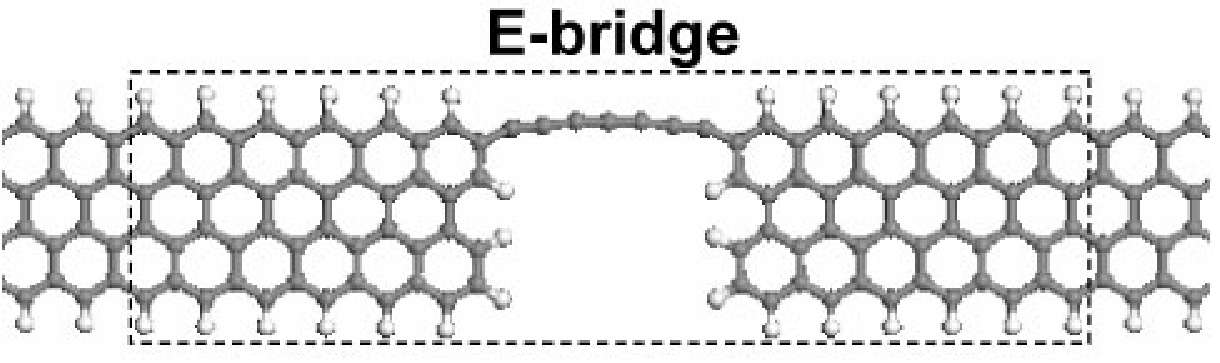} \\
(c) \includegraphics[width=7.3cm,clip]{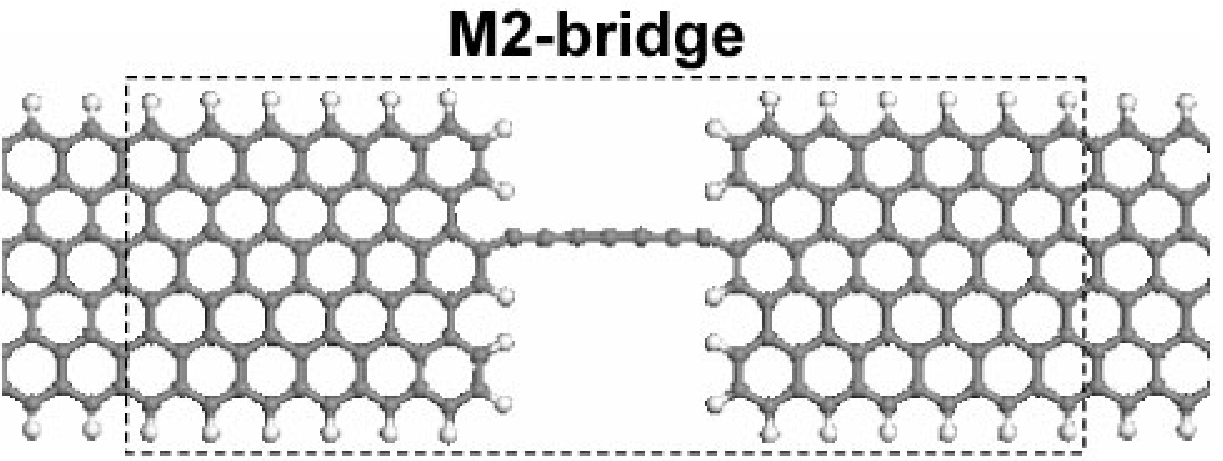} \\
(d) \includegraphics[width=7.3cm,clip]{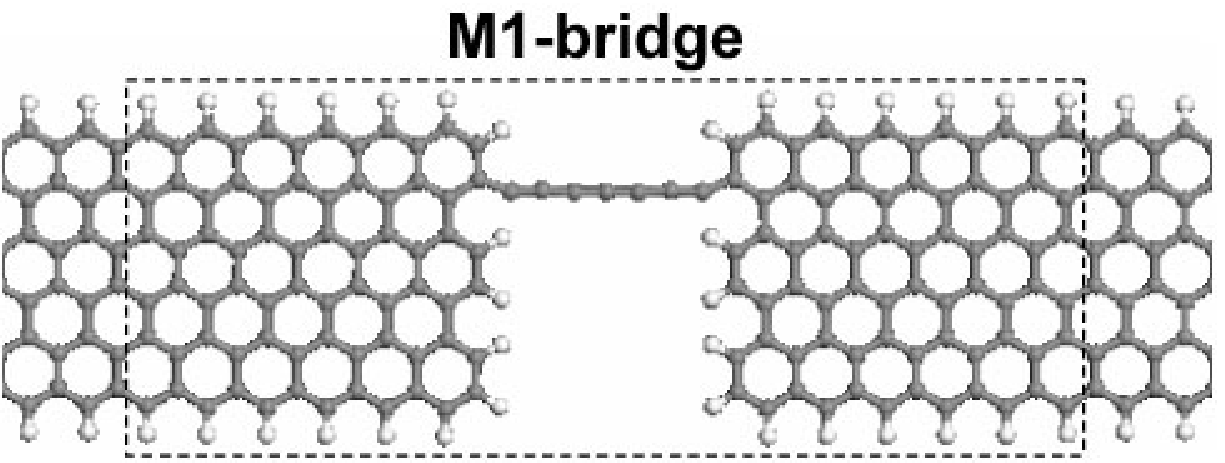} \\
(e) \includegraphics[width=7.3cm,clip]{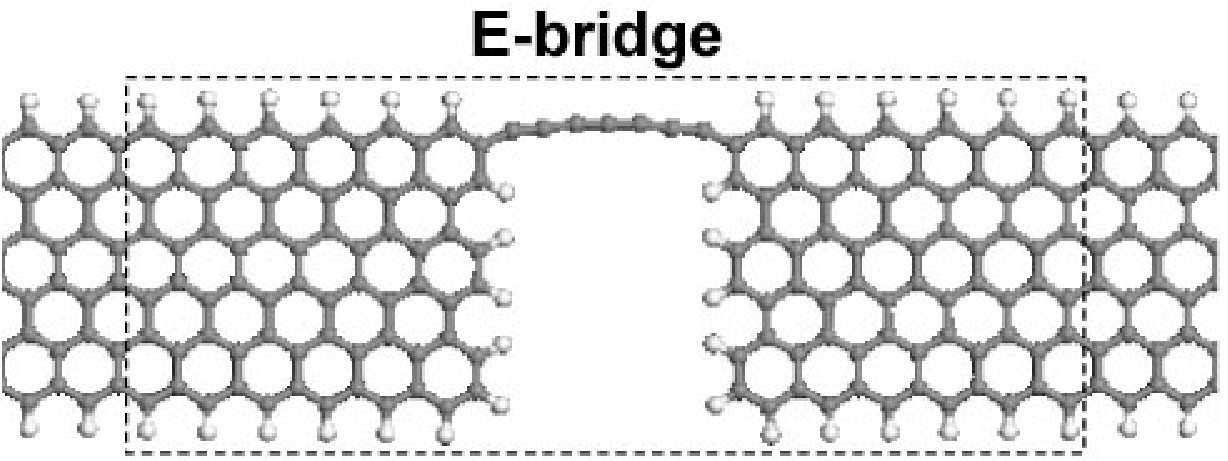}
\caption{\label{Fig_str}
Optimized atomic structures of the (4-ZGNR)-C$_7$-(4-ZGNR) and
(6-ZGNR)-C$_7$-(6-ZGNR) junctions with different bridge positions: (a),
(d) in the near middle (M1-bridge), (b), (e) at the edge (E-bridge), and (c) in
the right middle (M2-bridge). 
}
\end{figure}

\section{Computational details}

We consider two widths for the ZGNR leads, consisting of 4 and 6 zigzag carbon
chains, respectively (see Fig.~\ref{Fig_str}). A carbon chain with 7 atoms is
studied in details and a chain with 6 atoms is also calculated for showing the
even-odd behavior in the equilibrium conductance. 
For the junction with the 4-ZGNR leads, the carbon chain is connected at two
positions, one in the middle (M1-bridge) and the other at the edge (E-bridge).
For the junction with the 6-ZGNR leads, three bridge positions are
considered: in the right middle (M2-bridge), in the near middle (M1-bridge), and
at the edge (E-bridge), respectively, (see Fig.~\ref{Fig_str}). 

Each edge of the two ZGNR leads has two possible spin states, $\alpha$- or
$\beta$-spin. In this work, we consider two spin configurations for the leads.
In one configuration, both the left and right leads are in 
anti-parallel spin state with the top edge being $\alpha$-spin and the bottom
edge being $\beta$-spin (labeled by ($\alpha\beta$, $\alpha\beta$)), which is
the ground state of the narrow ZGNR lead. In the other, the two leads are in
parallel-spin state with all the four edges being $\alpha$-spin (labeled by
($\alpha\alpha$, $\alpha\alpha$)), which is possible under a magnetic field
and/or for wider ZGNR leads. Our calculation will show that the effect from the
width of the lead is quite small and therefore the narrow ZGNR leads considered
will also reflect the major behavior of wider ones. 

To investigate the electron transport through the carbon chain-ZGNR junctions we
adopt the NEGF-DFT approach \cite{ke2004electron,datta95} which combines the
NEGF formula for transport with {\it ab initio} DFT calculation for electronic
structure. In practice, the infinitely long 1-D system is divided into three
parts: left lead, right lead, and device region containing the carbon chain plus
enough ZGNR layers to accommodate the carbon chain-ZGNR interaction. The
self-consistent Kohn-Sham Hamiltonian of the device region and the self-energies
of two semi-infinite ZGNR leads are used to construct a single-particle Green's
function from which the transmission coefficient at any energy is calculated.
The conductance $G$ then follows from a Landauer-type relation. 
The computational techniques have been described in details previously
\cite{ke2004electron}. Briefly, for the DFT electronic structure calculation, we
use a numerical basis set to expand the wave function\cite{soler2002siesta}: A
double zeta plus polarization basis set (DZP) is adopted for all atomic species.
The local density approximation (LDA) \cite{PhysRevLett.45.566} is used for the
electron exchange and correlation and the optimized Troullier-Martins
pseudopotentials \cite{Troullier911993} are used for the atomic cores. The
atomic structure of the junctions including the carbon chain-ZGNR separation are
fully optimized by minimizing the atomic forces on the atoms to be smaller than
0.02 eV/{\AA}. 

\begin{figure}[tb]
\includegraphics[width=4.2cm,clip]{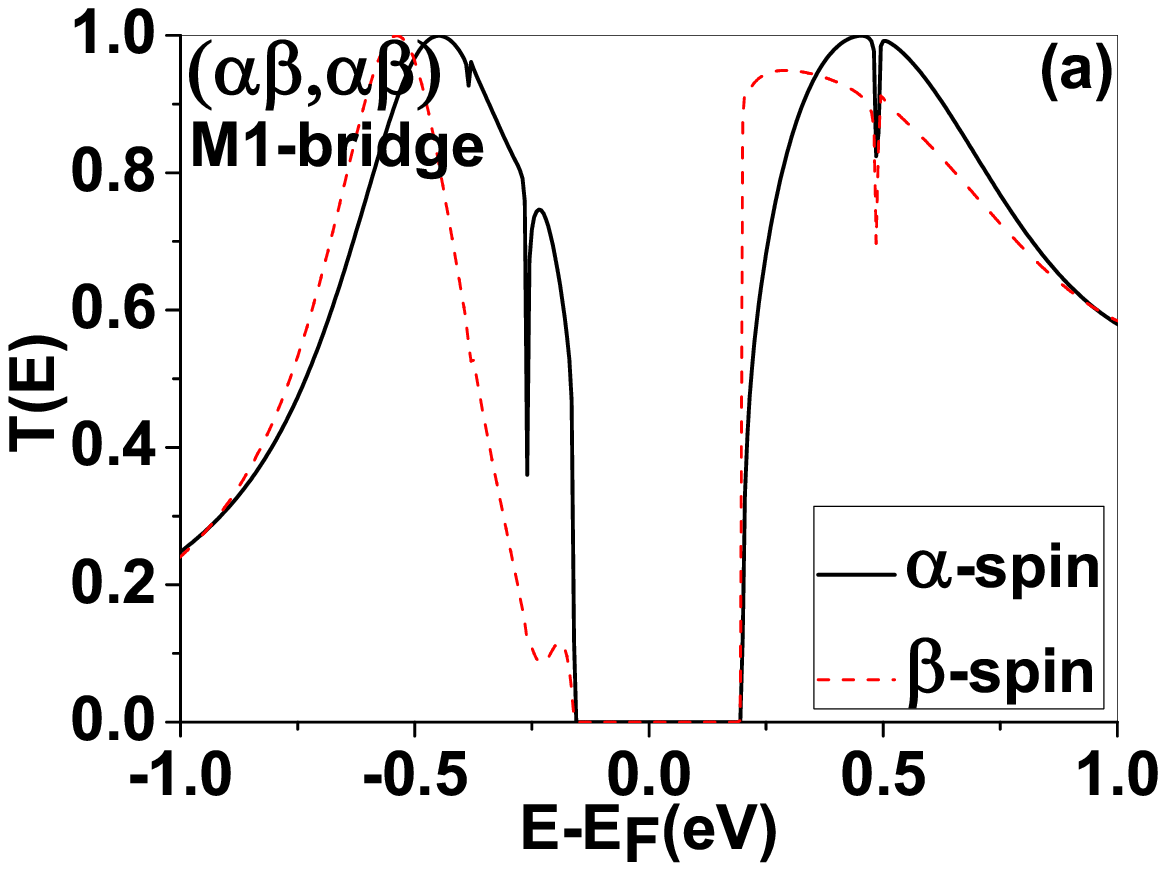}
\includegraphics[width=4.2cm,clip]{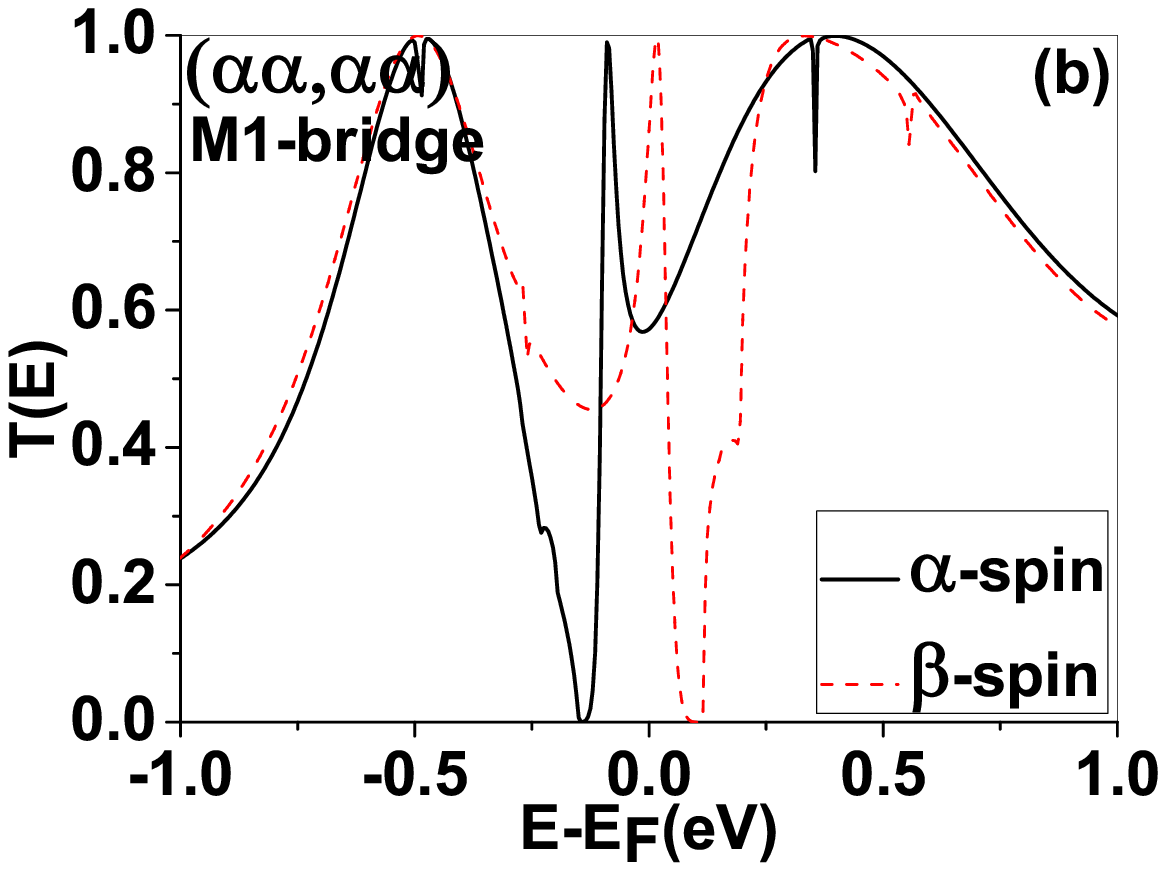}
\includegraphics[width=4.2cm,clip]{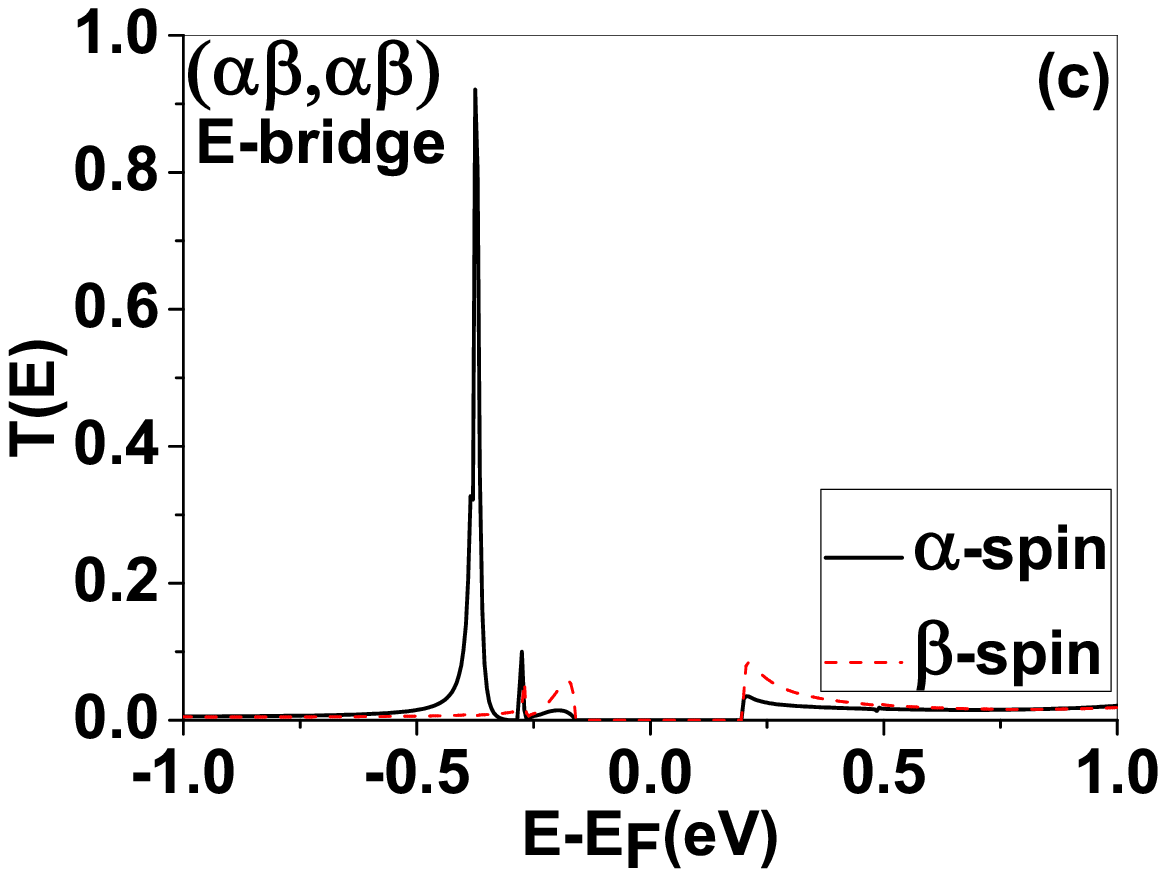}
\includegraphics[width=4.2cm,clip]{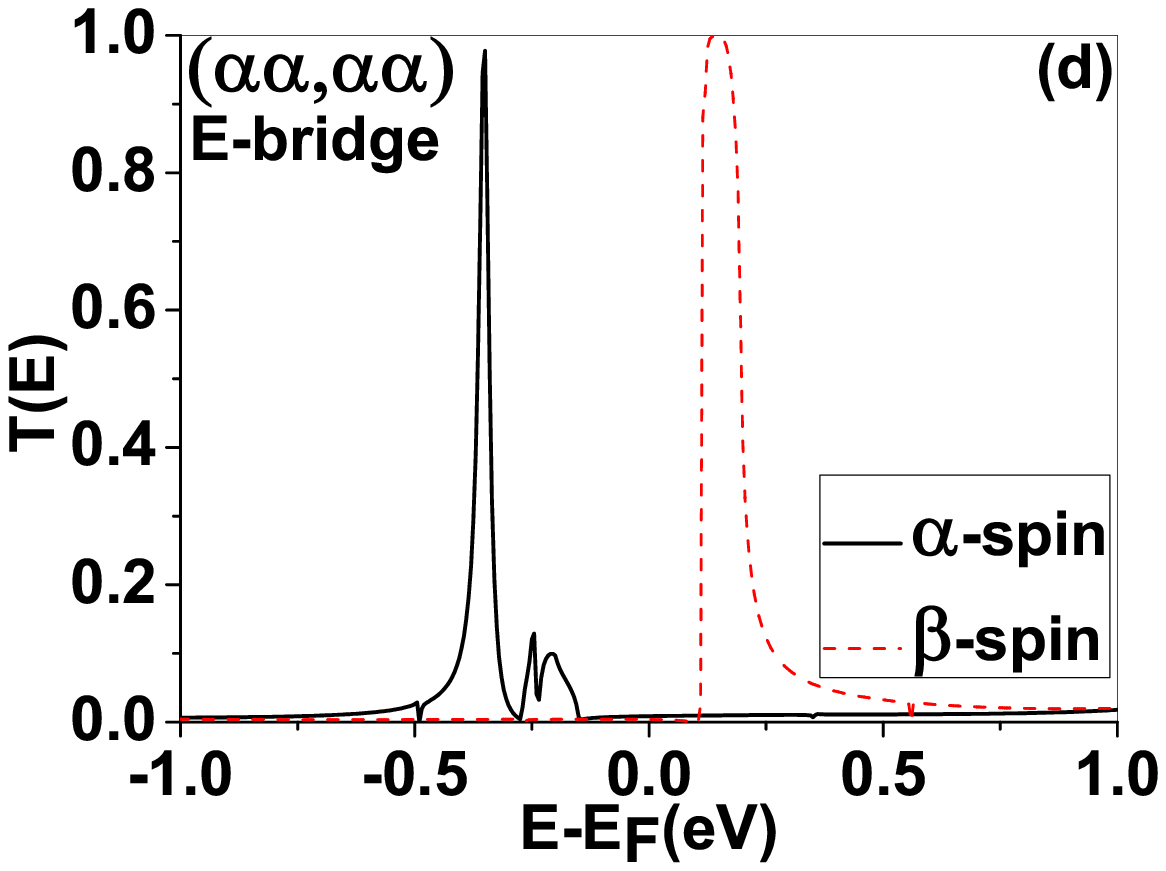}
\caption{\label{Fig_T(E)-4}
Transmission functions of the (4-ZGNR)-C$_7$-(4-ZGNR) junction.
The first panel: M1-bridge connection in
(a) ($\alpha\beta$, $\alpha\beta$) and (b) ($\alpha\alpha$, $\alpha\alpha$) spin configuration.
The second panel: E-bridge connection in (c) ($\alpha\beta$, $\alpha\beta$) and (d) ($\alpha\alpha$, $\alpha\alpha$)
spin configuration.
}
\end{figure}

\begin{figure}[tb]
\includegraphics[width=4.2cm,clip]{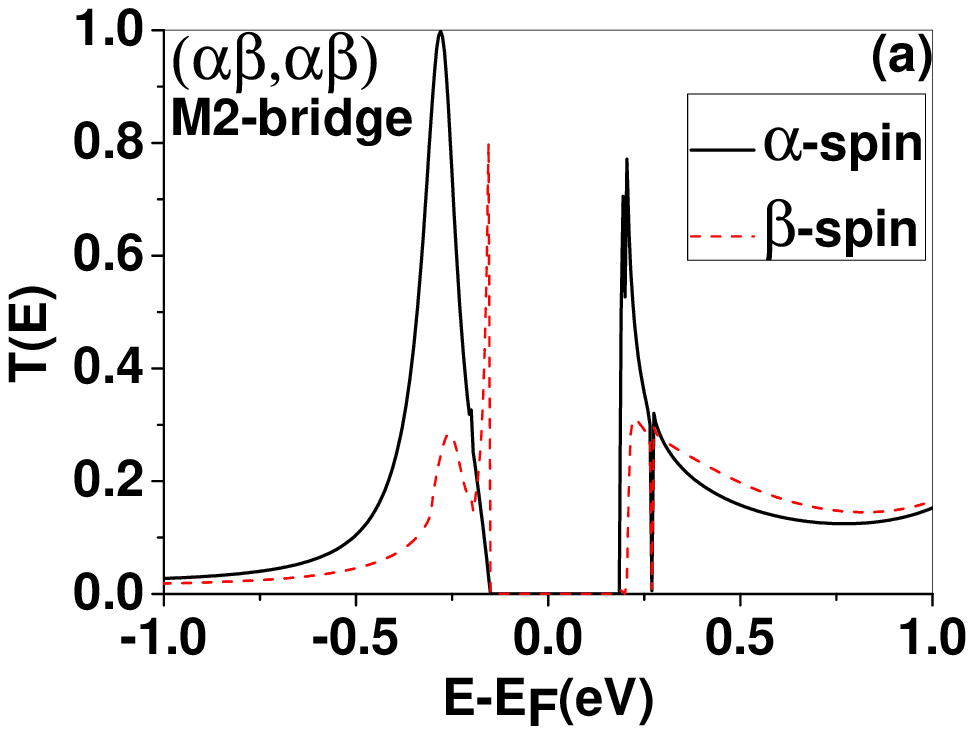}
\includegraphics[width=4.2cm,clip]{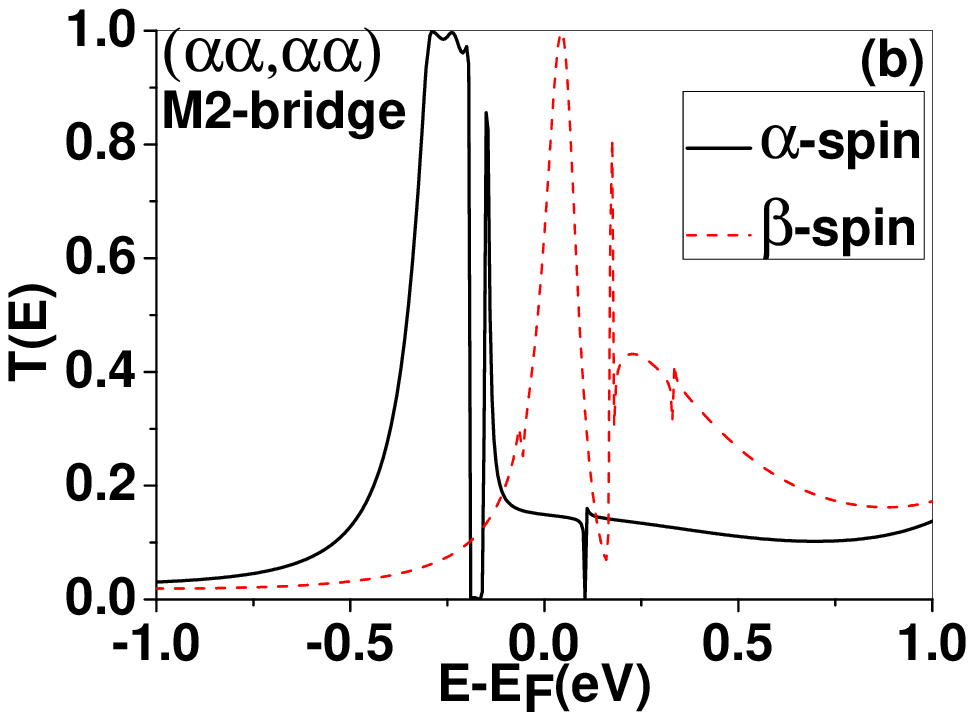}
\includegraphics[width=4.2cm,clip]{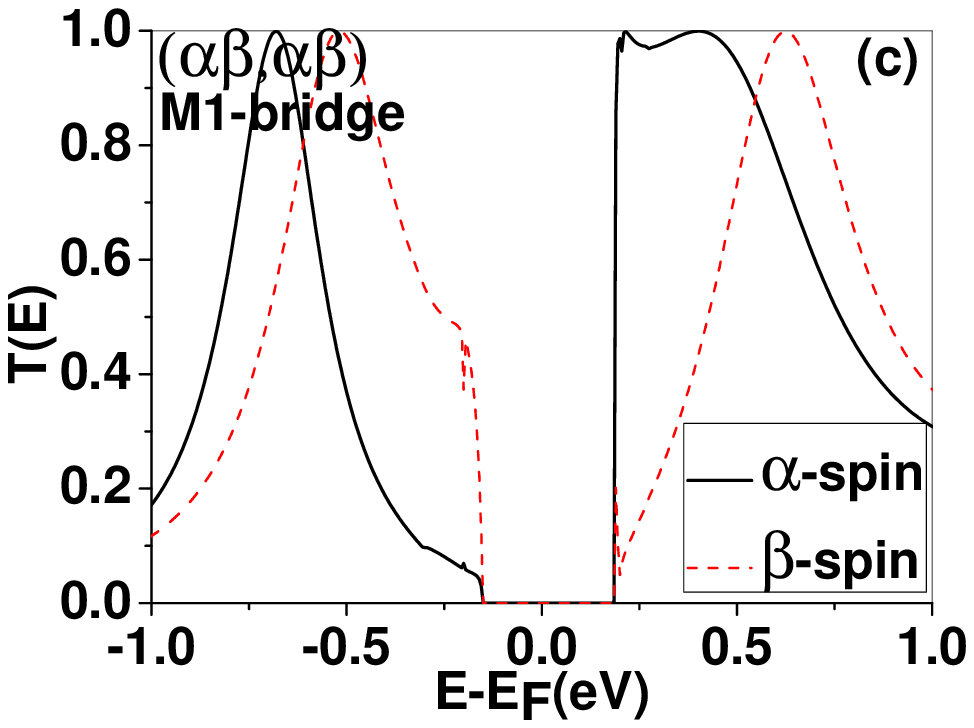}
\includegraphics[width=4.2cm,clip]{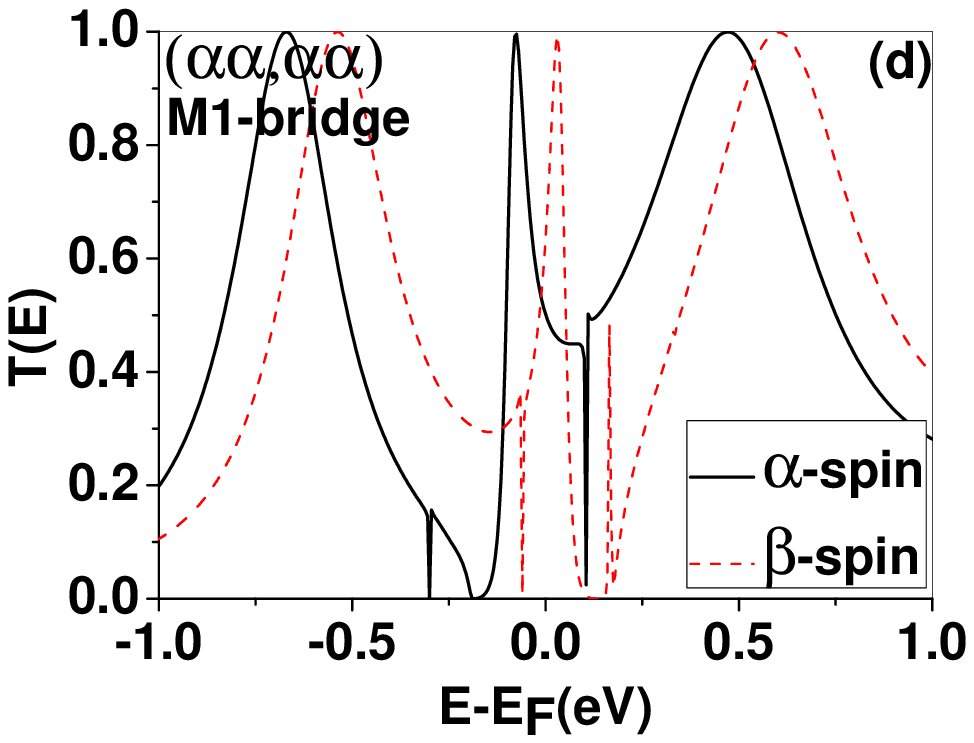}
\includegraphics[width=4.2cm,clip]{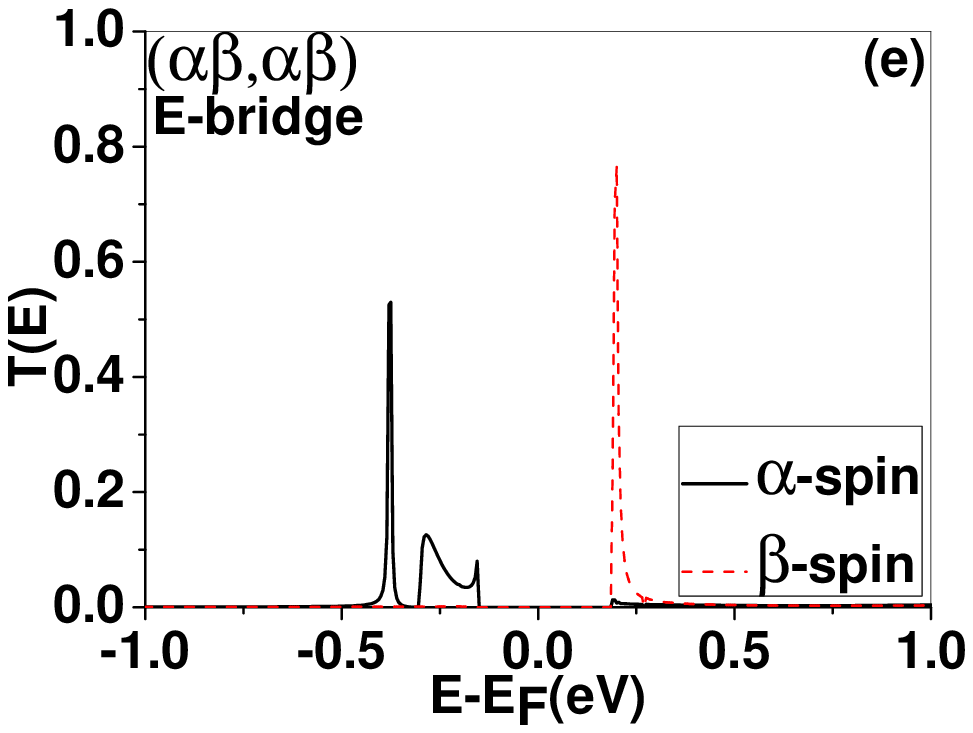}
\includegraphics[width=4.2cm,clip]{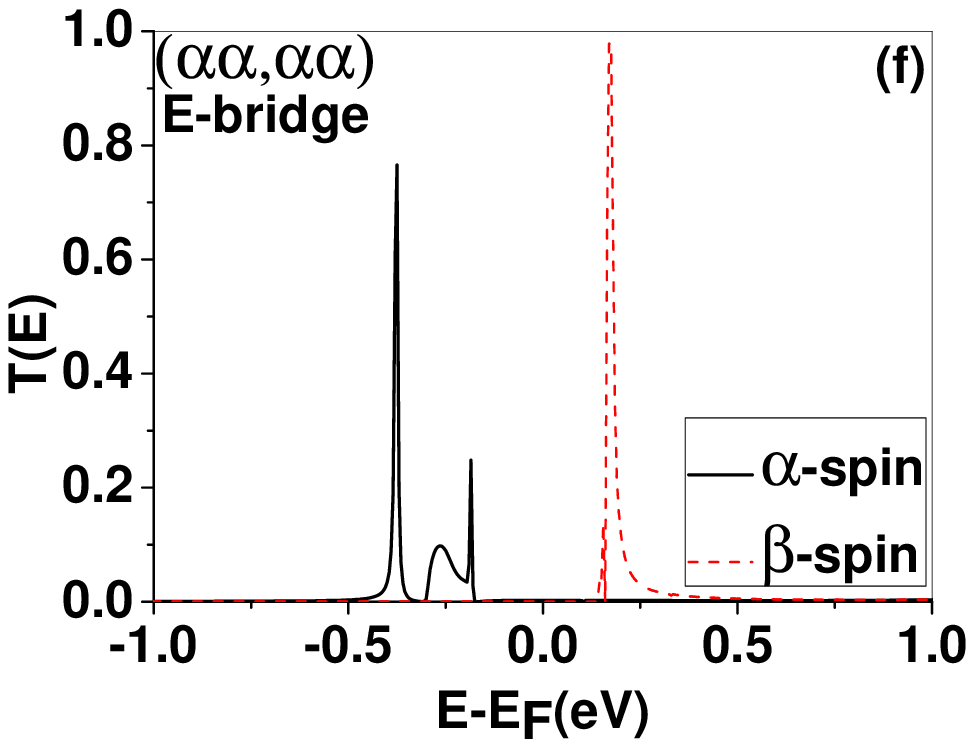}
\caption{\label{Fig_T(E)-6}
Transmission functions of the (6-ZGNR)-C$_7$-(6-ZGNR) junction.
The first panel: M2-bridge connection in
(a) ($\alpha\beta$, $\alpha\beta$) and (b) ($\alpha\alpha$, $\alpha\alpha$) spin configuration.
The second panel: M1-bridge connection in
(c) ($\alpha\beta$, $\alpha\beta$) and (d) ($\alpha\alpha$, $\alpha\alpha$) spin configuration.
The third panel: E-bridge connection in (e) ($\alpha\beta$, $\alpha\beta$) and (f) ($\alpha\alpha$, $\alpha\alpha$)
spin configuration.
}
\end{figure}

\section{Result and discussion}

\subsection{Effects of spin and bridge position}

\begin{figure}[tb]
\includegraphics[width=4.2cm,clip]{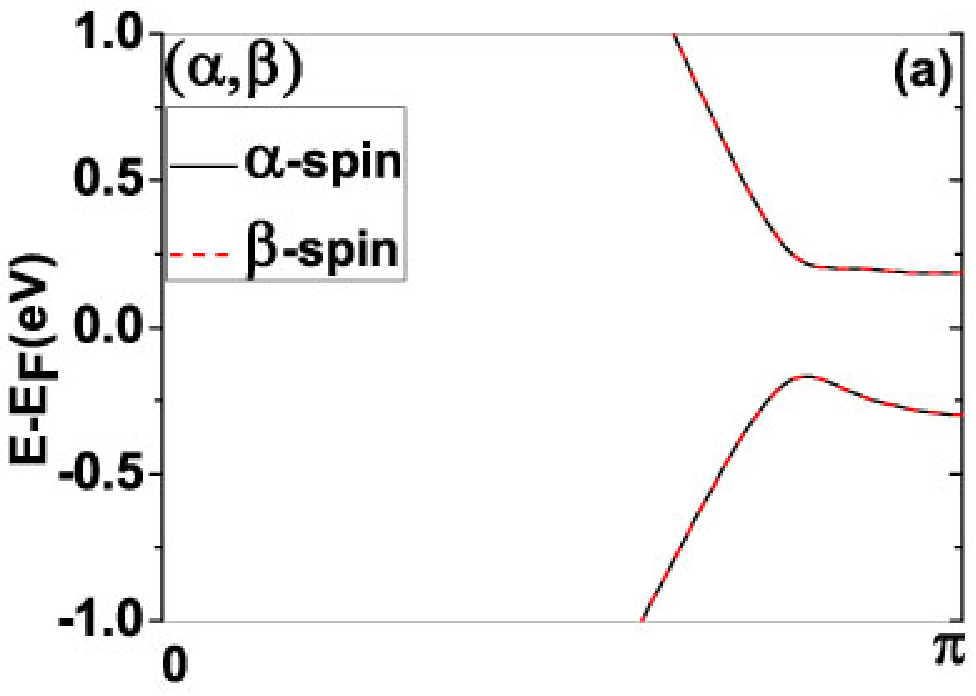}
\includegraphics[width=4.1cm,clip]{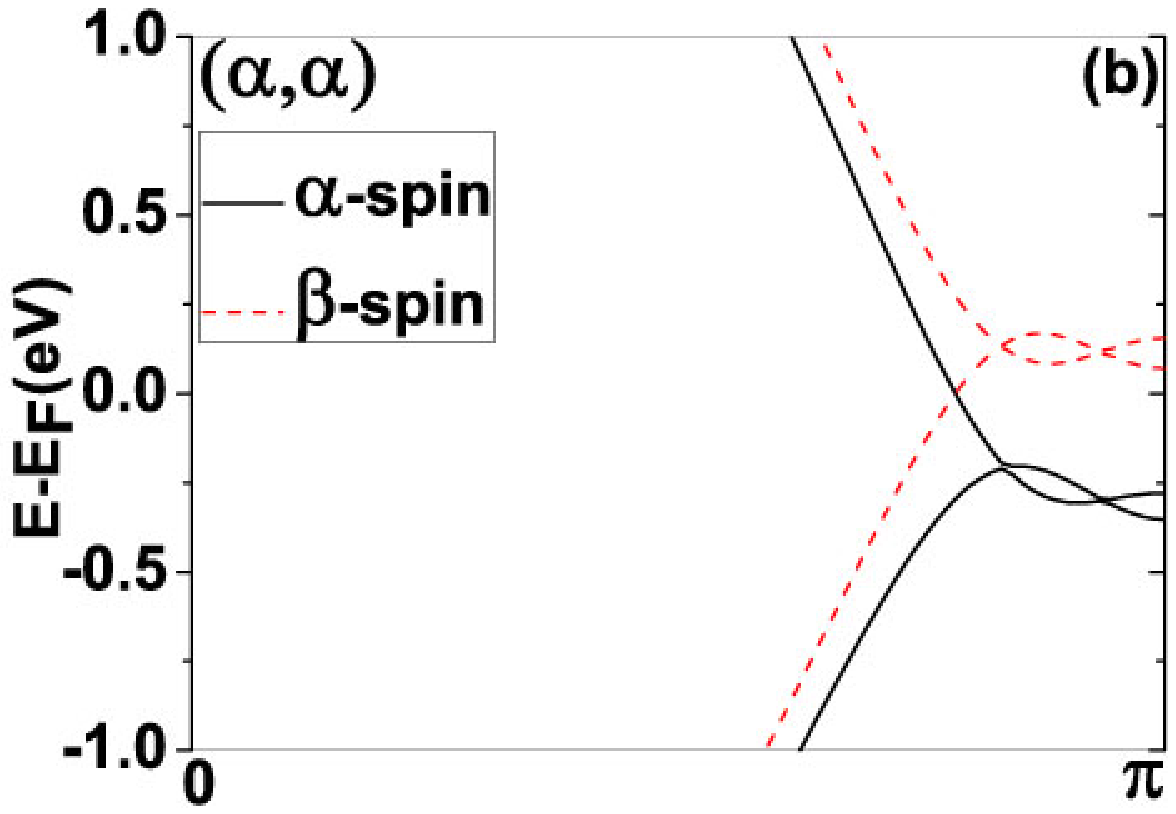}
\includegraphics[width=4.2cm,clip]{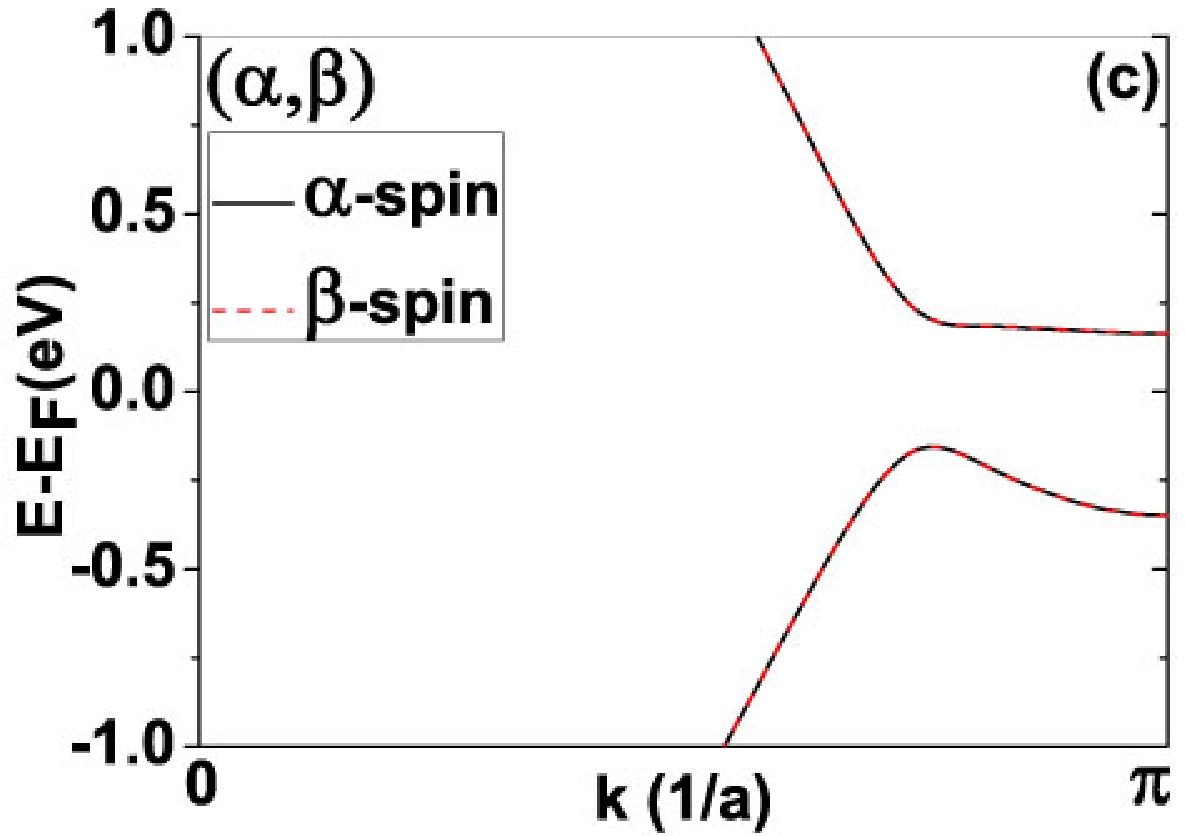}
\includegraphics[width=4.2cm,clip]{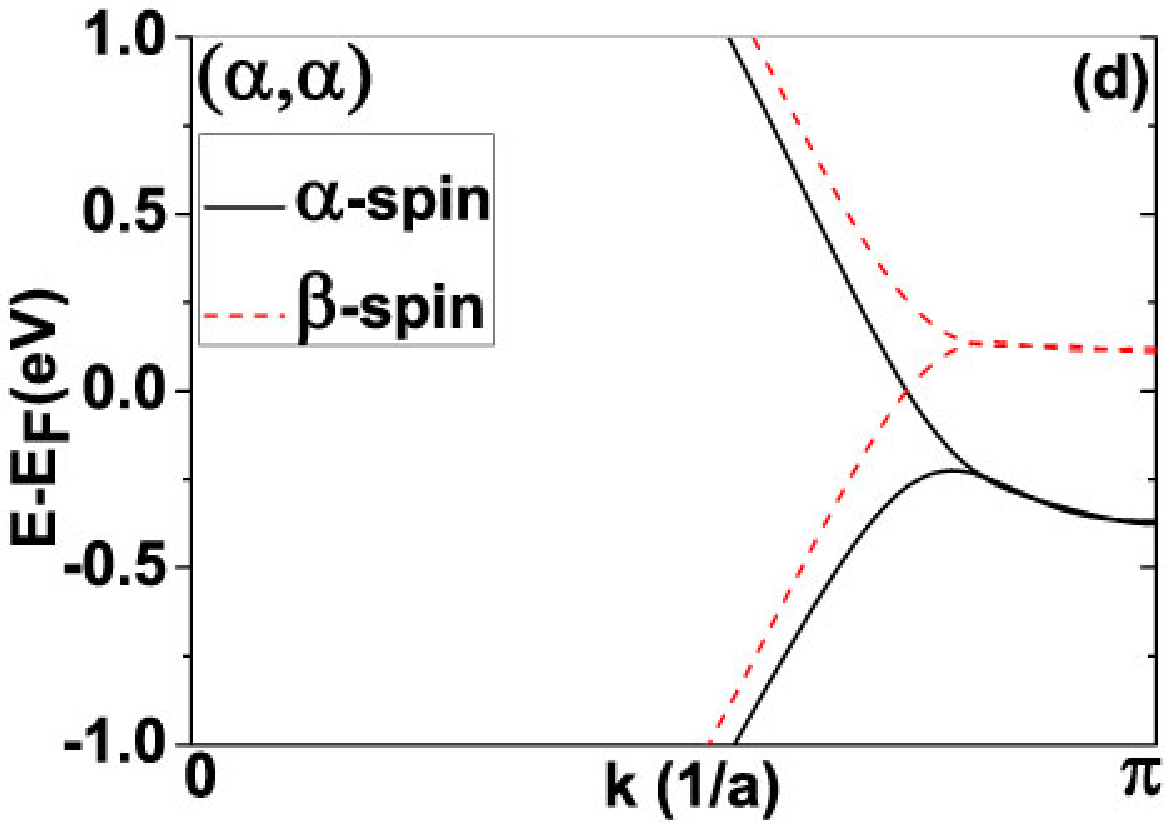}
\caption{\label{Fig_bands}
The first panel: Band structure of a 4-ZGNR in (a) ($\alpha$, $\beta$) and (b) ($\alpha$,
$\alpha$) spin configuration. The second panel: Band structure of a 6-ZGNR in (c)
($\alpha$, $\beta$) and (d) ($\alpha$, $\alpha$) spin configuration.
}
\end{figure}

The calculated transmission functions for the (4-ZGNR)-C$_7$-(4-ZGNR) and
(6-ZGNR)-C$_7$-(6-ZGNR) junctions are plotted in Figs. \ref{Fig_T(E)-4} and
\ref{Fig_T(E)-6}, respectively. The first thing to note is that the transmission
function depends very significantly on the spin configuration of the leads and
the position of the carbon chain bridge in the gap. 
When the bridge is positioned around the middle range of the gap (M2- and
M1-bridge) the transmission function is overall broad in the energy window [-1,
1] eV, indicating a strong coupling between the carbon chain and the ZGNR leads.
For the anti-parallel spin configuration ($\alpha\beta$, $\alpha\beta$) 
there is a small transport gap around the Fermi energy (Figs. \ref{Fig_T(E)-4}
(a) and \ref{Fig_T(E)-6} (a), (c)) while for the parallel spin confugration
($\alpha\alpha$, $\alpha\alpha$) there are sharp resonance peaks there (Figs.
\ref{Fig_T(E)-4} (b) and \ref{Fig_T(E)-6} (b), (d)). The transport gap in the
case of ($\alpha\beta$, $\alpha\beta$) is due to the band gap created by the
anti-parallel spins in the leads, as shown in Figs.~\ref{Fig_bands} (a) and (c).
Note that this band gap decreases slowly with the increasing width of the ribbon
and so is the transport gap: 0.37 and 0.35 eV for the 4-ZGNR and 6-ZGNR
junctions, respectively. 
In the case of ($\alpha\alpha$, $\alpha\alpha$), the parallel spin state causes
a band crossing at the Fermi energy as shown in Figs.~\ref{Fig_bands} (b) and
(d), providing an finite DOS around the Fermi energy in the leads. The coupling
of this finite DOS to the electronic states in the carbon chain gives rise to
the sharp resonance peaks around the Fermi energy (see futher discussion later).

On the other hand, when the carbon chain bridge is positioned at the edge of
the gap (E-bridge) the transmission function becomes very sharp peaks in the
energy window [-1, 1] eV (Figs. \ref{Fig_T(E)-4} (c), (d), and \ref{Fig_T(E)-6}
(e), (f) ), indicating an overall weak chain-lead coupling. Furthermore, around
the Fermi energy the $\alpha$- and $\beta$-spin peaks are now separated largely
from each other, making the transport at the Fermi energy is also blocked for the
($\alpha\alpha$, $\alpha\alpha$) spin configuration (Figs. \ref{Fig_T(E)-4} (d)
and \ref{Fig_T(E)-6} (f) ). 

Being consistent with the previous calculation adopting wider ZGNR leads
\cite{zanolli2010quantum}, our calculation shows that the C$_7$ chain is spin-polarized by itself
\cite{spin-polarization}. This leads to the spin-polarized transport even when the carbon
chain is connected in the right middle (i.e., M2-bridge, see Fig.\ref{Fig_T(E)-6} (b)).
However, the present calculation further shows that the resulting spin-polarized
transport is significantly modulated by the bridge position.
For example, in the case of ($\alpha\alpha$, $\alpha\alpha$) the splitting
between the $\alpha$- and $\beta$-spin peaks and also the spin polarization
ratio around the Fermi energy varies largely with the bridge position 
(see Fig.\ref{Fig_T(E)-6} (b) vs (d)) due to the interaction with the edge
states (see further discussion later).

As for the effect from the width of the lead, one can see that for the same
M1-bridge connection the overall effect in the transmission function is not
significant and does not change the qualitative result when the width of the
leads is increased from 4 to 6 zigzag chains: Fig.~\ref{Fig_T(E)-4} (a) vs
Fig.~\ref{Fig_T(E)-6} (c) for ($\alpha\beta$, $\alpha\beta$), and
Fig.~\ref{Fig_T(E)-4} (b) vs Fig.~\ref{Fig_T(E)-6} (d) for ($\alpha\alpha$,
$\alpha\alpha$). Therefore, the results from the present calculation can be
expected to be still qualitatively valid for junctions with wider ZGNR leads. 

\begin{figure}[bt]
\includegraphics[width=4.2cm,clip]{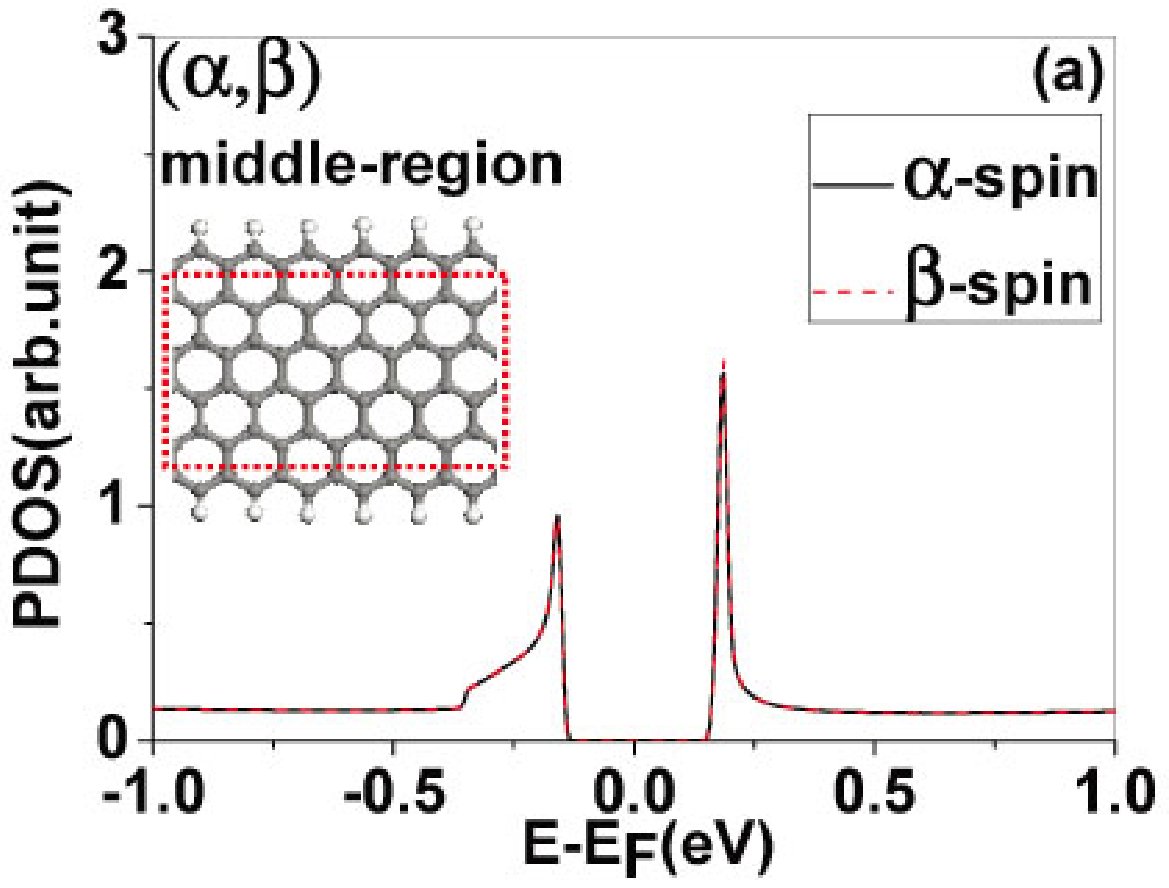}
\includegraphics[width=4.2cm,clip]{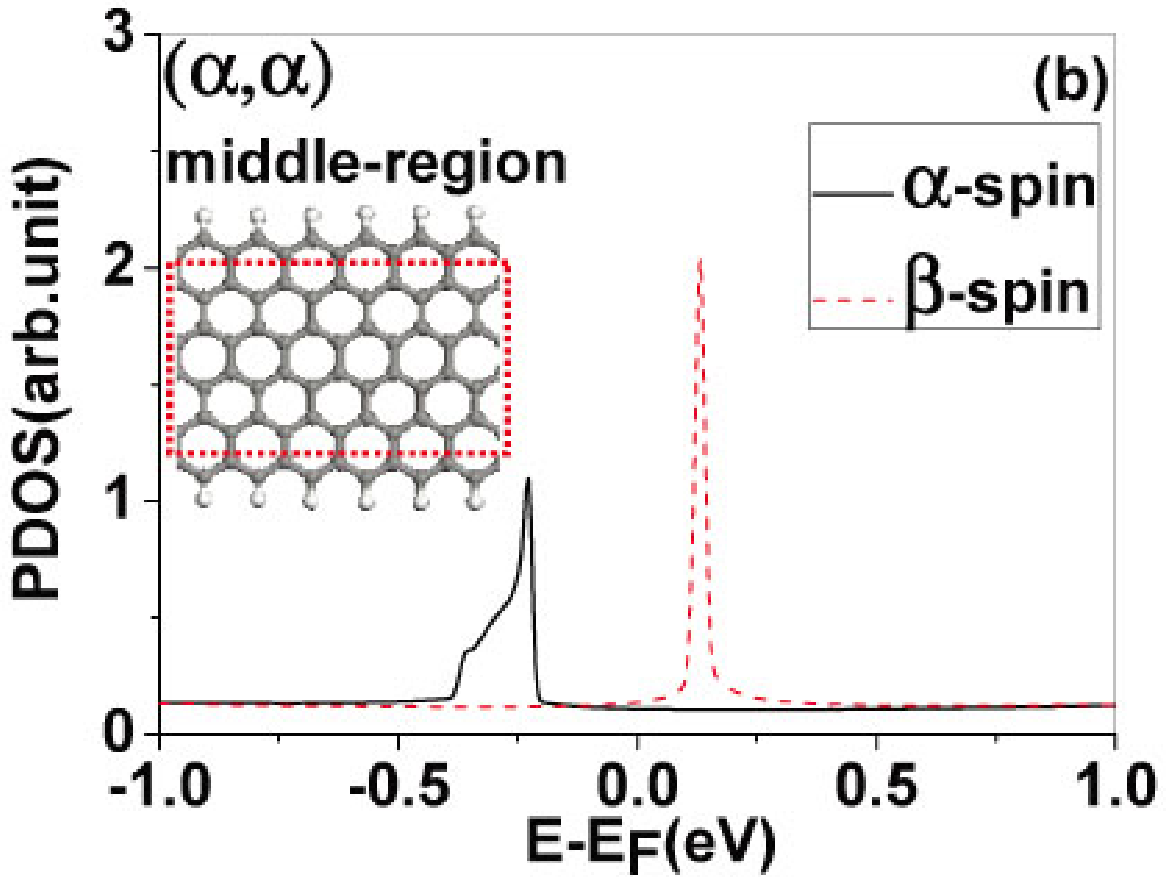}
\includegraphics[width=4.2cm,clip]{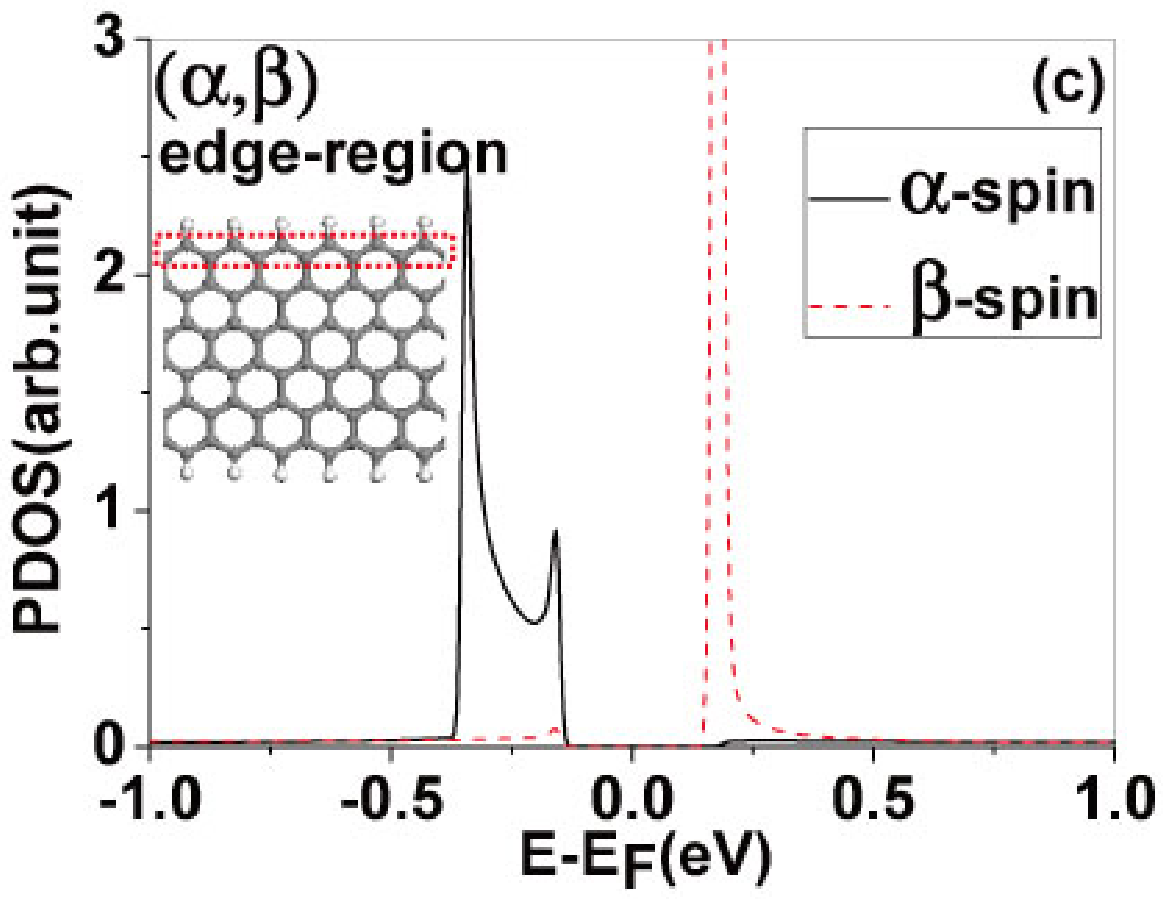}
\includegraphics[width=4.2cm,clip]{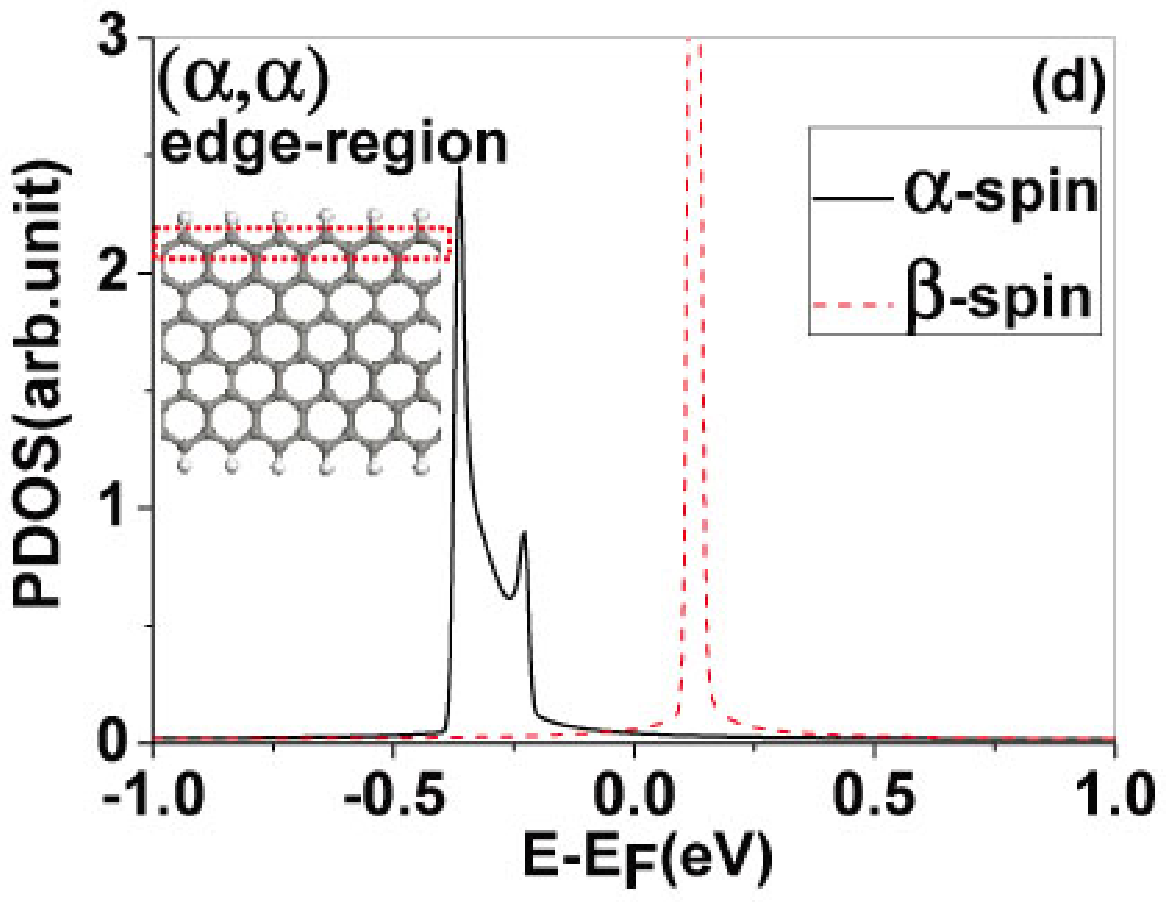}
\includegraphics[width=4.2cm,clip]{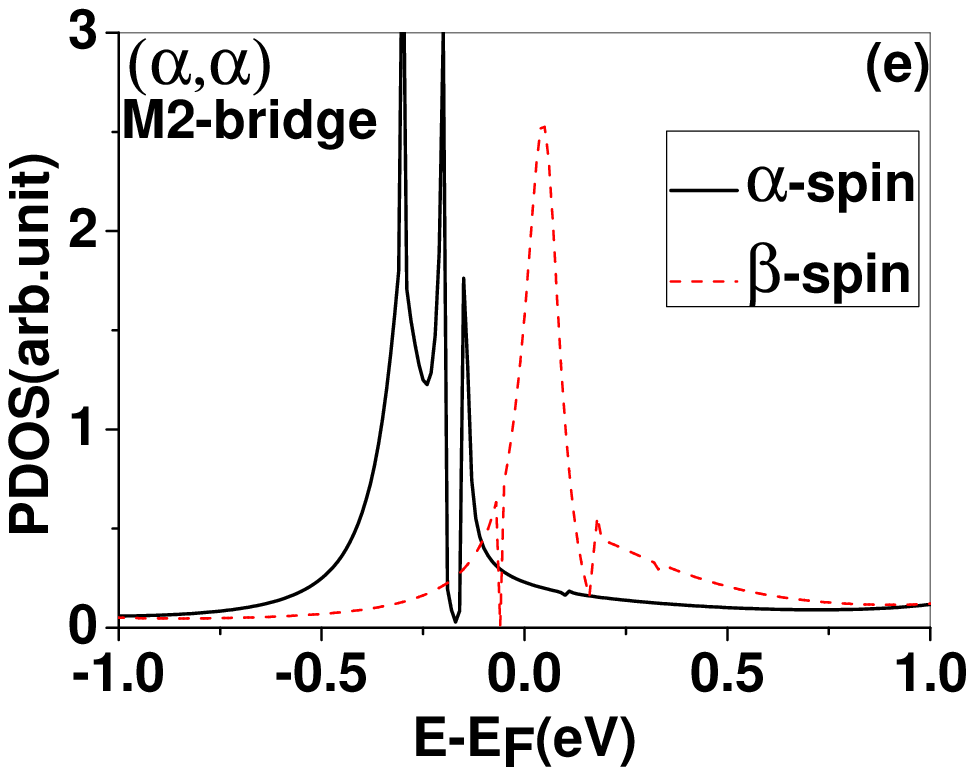}
\includegraphics[width=4.2cm,clip]{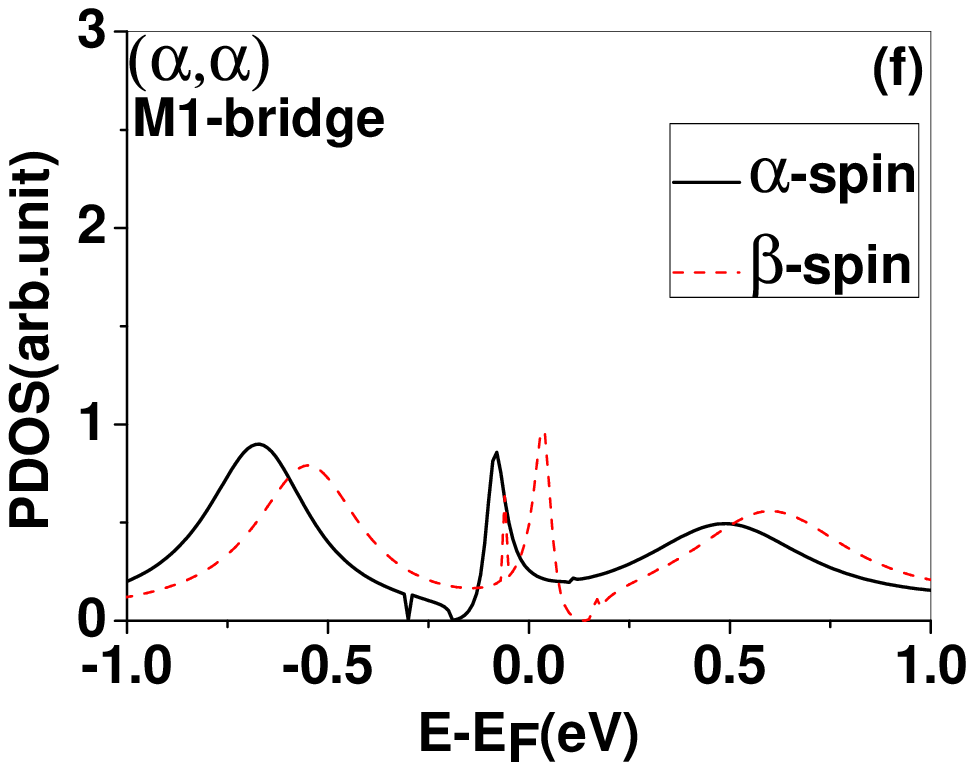}
\caption{\label{Fig_PDOS}
The first panel: PDOS projected on the bulk region of a 6-ZGNR for
(a) ($\alpha$, $\beta$) and (b) ($\alpha$, $\alpha$) spin configuration.
The second panel: PDOS projected on the edge region of a 6-ZGNR for
(c) ($\alpha$, $\beta$) and (d) ($\alpha$, $\alpha$) spin configuration.
The third panel:
PDOS projected on the C$_7$ chain in the (6-ZGNR)-C$_7$-(6-ZGNR) junction
with (e) M2-bridge and (f) M1-bridge connection in the
($\alpha\alpha$, $\alpha\alpha$) spin configuration.
}
\end{figure}

\begin{figure}[tb]
(a) \includegraphics[width=7cm,clip]{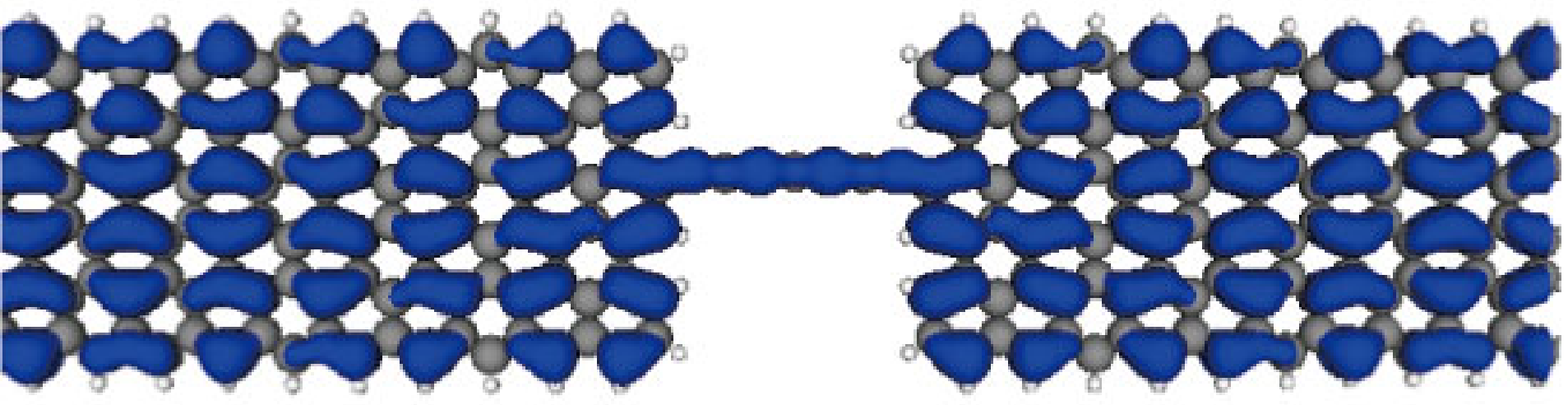} \\
(b) \includegraphics[width=7cm,clip]{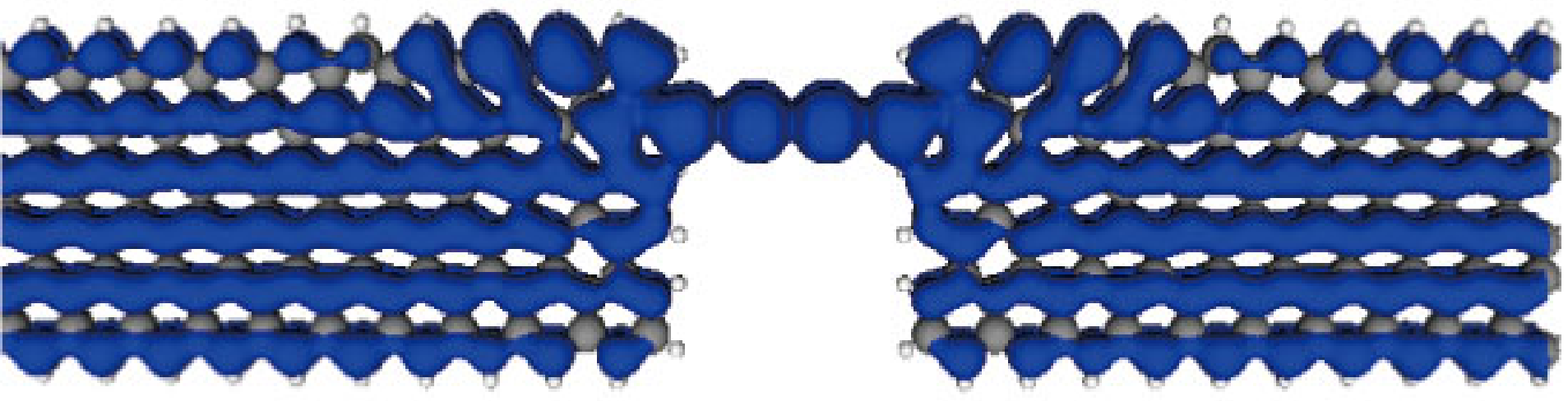}
\caption{\label{Fig_ldos-a}
(a) and (b): LDOS of the $\alpha$-spin component around -0.68 eV in Fig.\ref{Fig_T(E)-6} (a) and (c), respectively.
}
\end{figure}

\begin{figure}[tb]
(a)\includegraphics[width=7cm,clip]{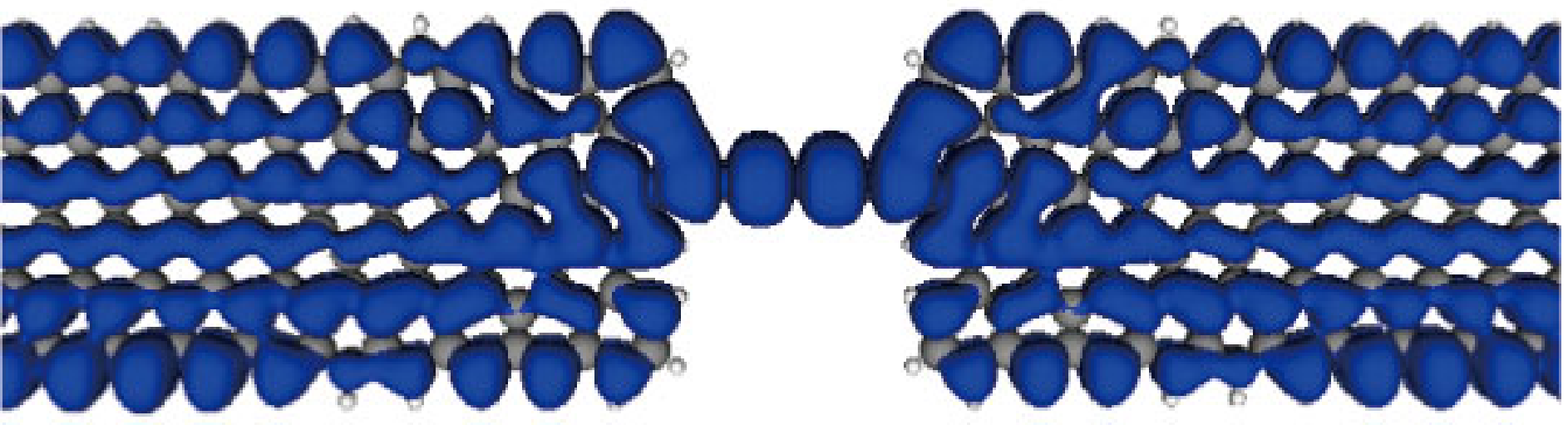} \\
(b) \includegraphics[width=7cm,clip]{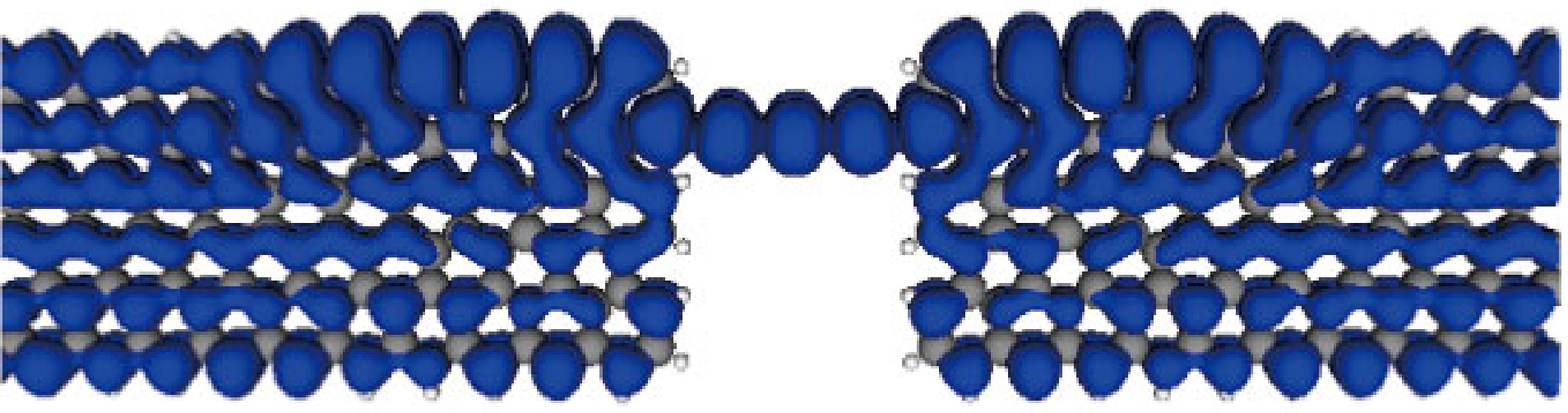} \\
(c) \includegraphics[width=7cm,clip]{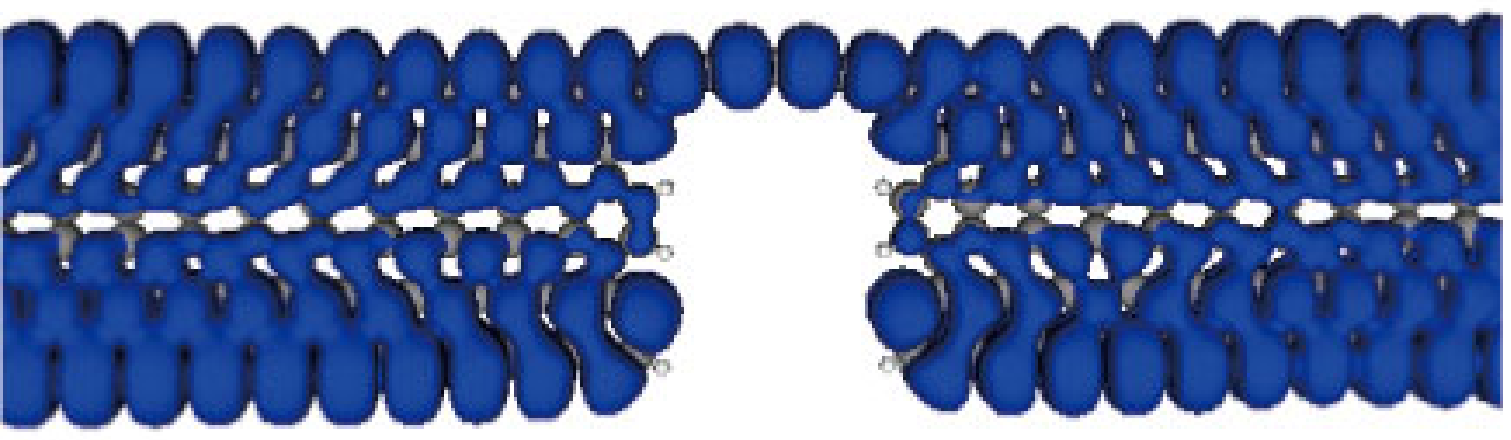}
\caption{\label{Fig_ldosB}
(a), (b), (c): LDOS of the $\beta$-spin peaks around the Fermi energy (0.05, 0.03, and 0.17 eV, respectively)
in Fig.\ref{Fig_T(E)-6} (b), (d), and (f), respectively.
}
\end{figure}

\subsection{Analyses}

Physically, the complicated spin-dependent transport behavior and the very
significant effect from the bridge position is related to the spin-polarized
edge states and their spatial distribution in the leads. To understand the
results and provide an insight into the physics underlying, we study the
spin-polarized projected density of states (PDOS) for the leads and the local
density of states (LDOS) for the characteristic peaks in the transmission
function of the junctions. 

In Fig.~\ref{Fig_PDOS} we show the spin-polarized PDOS for the edge and bulk
regions of an infinite long 6-ZGNR. In the bulk region of the ribbon
(Figs.~\ref{Fig_PDOS} (a) and (b) ) the bulk states provide a finite DOS in the
whole energy window except for that around the Fermi energy where a gap appears
for ($\alpha$, $\beta$) because of the band gap shown in Fig.~\ref{Fig_bands}
(c), while a finite DOS still exists for ($\alpha$, $\alpha$) due to the band
crossing shown in Fig.~\ref{Fig_bands} (d). When the bridge is positioned in the
middle region of the gap the finite DOS apart from the Fermi energy couples with
the states in the bridge, leading to the very broad transmission function shown
in Figs.~\ref{Fig_T(E)-6} (a) - (d). To show this more clearly, we plot in
Fig.~\ref{Fig_ldos-a} the LDOS for the energy around -0.68 eV in
Figs.~\ref{Fig_T(E)-6} (a) and (c), respectively. It can be seen that in both
cases the LDOS spreads out through the whole junction and distributes quite
evenly in the leads, indicating that it is associated with the bulk states. In
the case of ($\alpha$, $\alpha$) the finite DOS around the Fermi energy couples
with the spin-polarized states in the bridge (their PDOS is plotted in Figs.~\ref{Fig_PDOS} (e)
and (f)) and gives rise to the resonance peaks around the Fermi energy in the
transmission function (Figs.~\ref{Fig_T(E)-6} (b) and (d)). To show the nature
of these resonance peaks we plot their LDOS for the $\beta$-spin component 
in Figs.~\ref{Fig_ldosB} (a) and (b) for the M2- and M1-bridge connections,
respectively. One can see that when the bridge is positioned in the right middle
(M2-bridge) the LDOS has large contribution from the bridge region and is
distributed evenly throughout the leads, indicating that it is a result of the
coupling between the states in the bridge and the bulk states in the leads. When
the bridge is closer to the edge (M1-bridge) now the LDOS has also large
contribution from the edge states within the scattering region, indicating 
an interaction between the spin-polarized bridge states and the edge states,
which modulates the spin-polarized transport behavior of the junction around the
Fermi energy (see Fig.\ref{Fig_T(E)-6} (b) vs (d)). 

On the other hand, for the zigzag chain of the edge (Figs.~\ref{Fig_PDOS} (c)
and (d) ) the PDOS reflects essentially the edge state which has two sharp peaks
beside the Fermi energy, i.e., the occupied $\alpha$-spin and unoccupied
$\beta$-spin states, respectively, placing a near-zero DOS elsewhere.
Consequently, when the bridge is positioned at the edge it mainly couples with
these edge states, leading to the transmission function with only very sharp
peaks corresponding to the edge states, as shown in Figs.~\ref{Fig_T(E)-6} (e)
and (f). The LDOS of the $\beta$-spin peak in Fig.~\ref{Fig_T(E)-6} (f) plotted
in Figs.~\ref{Fig_ldosB} (c) shows that the transport is indeed along the two
edges. In this case, even for the ($\alpha\alpha$, $\alpha\alpha$) spin
configuration the transport around the Fermi energy is also nearly blocked. 
Note that unlike the transport gap created by the anti-parallel spins in
($\alpha\beta$, $\alpha\beta$) case, this transport gap between the $\alpha$-
and $\beta$-spin conponents actually increases from the 4-ZGNR to the 6-ZGNR
junction (0.26 and 0.33eV, respectively). This is because the $\alpha-\alpha$
spin interaction in the narrow 4-ZGNR lead induces extra dispersions of the edge
states, as shown in Fig.~\ref{Fig_bands} (b) vs (d). 

The broad and very narrow peaks in the transmission function reflect the overall
coupling strength between the carbon chain and the ZGNR leads. It is interesting
to note that the best coupling is given by neither the bridge at the edge
(E-bridge) nor the bridge right in the middle (M2-bridge) but the one positioned
in between (M1-bridge). This is evident in the LDOS shown in
Fig.~\ref{Fig_ldos-a} where the M1-bridge gives a remarkably larger LDOS around
the bridge and contact region than the M2-bridge does. The result is a much
larger transmission coefficient around -0.68 eV in Fig.~\ref{Fig_T(E)-6} (c)
than that in Fig.~\ref{Fig_T(E)-6} (a). 

\begin{figure}[tb]
\includegraphics[width=5.0cm,clip]{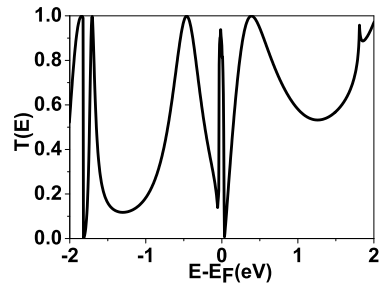}
\caption{\label{Fig_T(E)-4-nospin}
Transmission function of the spin-unpolarized (4-ZGNR)-C$_7$-(4-ZGNR) junction.
}
\end{figure}

\subsection{Even-odd behavior}

Finally, we would like to discuss the even-odd behavior in the equilibrium
conductance found previously in the spin-unpolarized calculation
\cite{shen2010electron}. This calculation shows that short carbon atomic chains
with an odd number of atom will give a significantly larger equilibirum
conductance than those with an even number of atom. According to the present
calculation, however, the spin freedom must be taken into account in reaching
this conclusion. To have a direct comparison, we also perform spin-unpolarized
calculation for the narrow 4-ZGNR junction with the M1-bridge and plot the
result in Fig.~\ref{Fig_T(E)-4-nospin}, which is basically the same as obtained
in Ref.~\cite{shen2010electron}. The very sharp peak at the Fermi energy is the
origination of the even-odd oscillation in conductance, which was ascribed to
the edge states \cite{shen2010electron}. The present spin-polarized result in
Fig.~\ref{Fig_T(E)-4} shows that this peak around the Fermi energy only exists
in the case of ($\alpha\alpha$, $\alpha\alpha$) with the M1-bridge connection
(see Fig.~\ref{Fig_T(E)-4} (b)) but now having a small splitting between the
$\alpha$- and $\beta$-spin components. 
In the case of ($\alpha\beta$, $\alpha\beta$) the equilibrium conductance is
zero simply because the band gap created by the anti-parallel spins in the leads
(see Fig.~\ref{Fig_T(E)-4} (a)). Additionally, for the E-bridge connection even
in the case of ($\alpha\alpha$, $\alpha\alpha$) the equilibrium conductance is
also nearly zero because of the large splitting between the $\alpha$- and
$\beta$-spin components (see Fig.~\ref{Fig_T(E)-4} (d)). For the wider
(6-ZGNR)-C$_7$-(6-ZGNR) junction, the result is similar -- the sharp peaks
around the Fermi energy only appear in the case of ($\alpha\alpha$,
$\alpha\alpha$) and only when the bridge is positioned around the middle region
(see Fig.~\ref{Fig_T(E)-6}). 

The present spin-polarized calcuation shows that, as can be seen in
Figs.~\ref{Fig_ldosB} (a) and (b), the large resonance peaks at the Fermi energy
is not originated from the edge states, but 
comes from the coupling between the finite DOS in the leads due to the band
crossing and the states in the bridge whose PDOS is given in
Figs.~\ref{Fig_PDOS} (e) and (f) showing a large DOS around the Fermi 
energy for the C$_7$ chain. 
Since in this case the coupling to the leads is determined by the states in the
carbon chain, the resulting equilibrium conductance will be sensitive to its
electronic state which may be affected significantly by the number of atoms in
the chain. 
As was found in Ref.~\cite{shen2010electron}, an even or odd number of atoms in
the carbon chain will result in different C-C bond-length distribution due to
the Peils transition effect and therefore gives quite different electronic
states.

\begin{figure}[tb]
\includegraphics[width=4.2cm,clip]{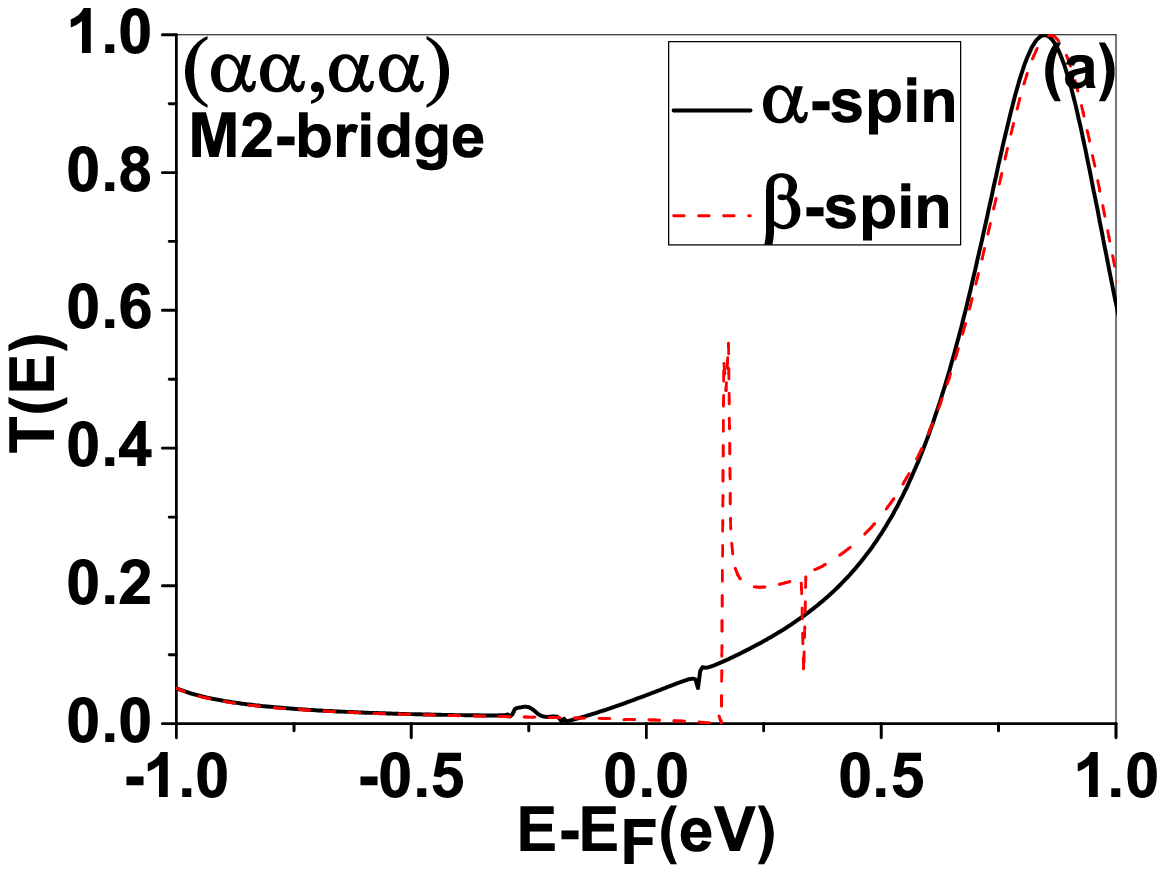}
\includegraphics[width=4.2cm,clip]{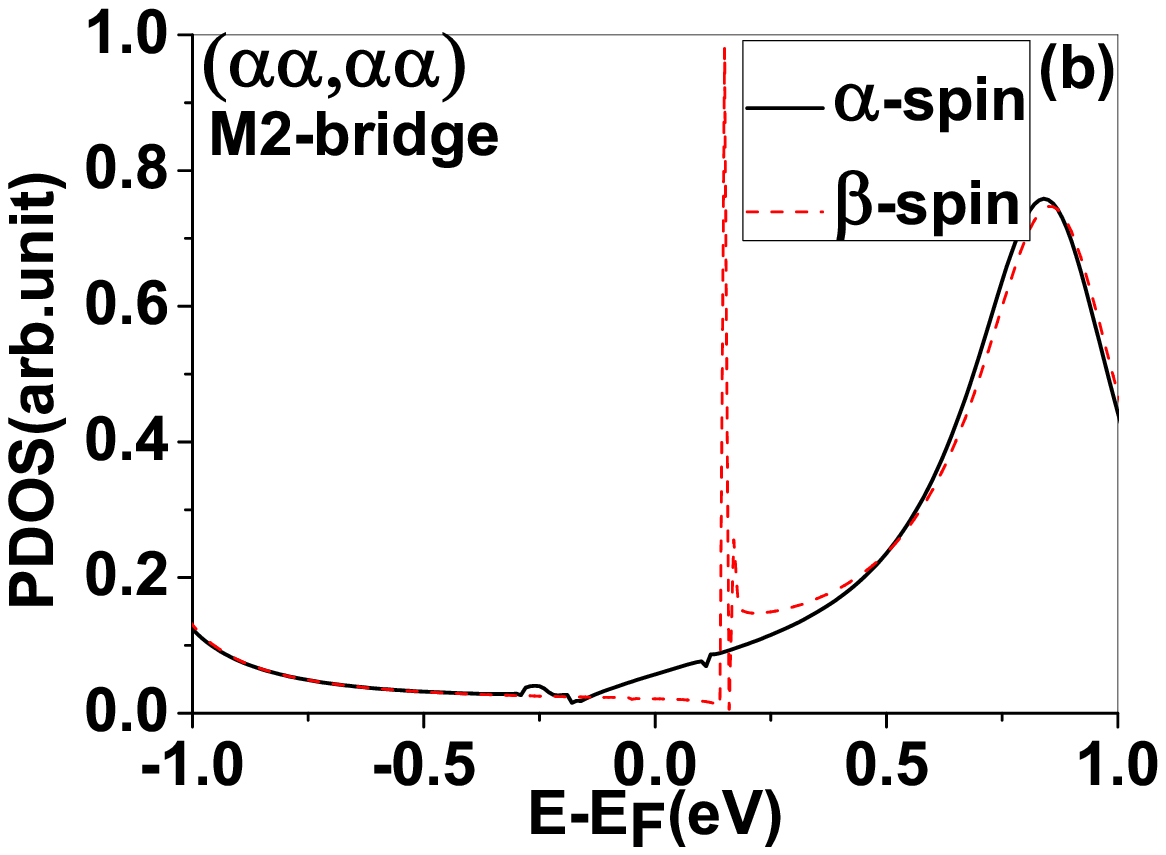}
\caption{\label{AA-6-wider-M1}
(a) Transmission function of the (6-ZGNR)-C$_6$-(6-ZGNR) junction in the ($\alpha\alpha$, $\alpha\alpha$)
spin configuration and with the M2-bridge connection. (b) PDOS projected on the C$_6$ chain of the junction.
}
\end{figure}

In order to make this issue more clear, we calculate the transimission function
of a (6-ZGNR)-C$_6$-(6-ZGNR) junction with the M2-bridge connection for the
($\alpha\alpha$, $\alpha\alpha$) spin configuration. The result and the PDOS
projected on the C$_6$ chain are plotted in Figs.~\ref{AA-6-wider-M1} (a) and
(b), respectively. The small DOS at Fermi energy for the  C$_6$ chain shows the
the coupling between the lead and the carbon chain is very weak, resulting a
much smaller equilibrium conductance compared with the  C$_7$ junction. The
consequence is an even-odd oscillation in the equilibrium conductance in terms
of the number of atoms in the carbon chain. 

\section{Summary}

In summary, motivated by recent experiments of successfully carving out stable
carbon atomic chains from graphene, we have studied the spin-dependent electron
transport through carbon atomic chains connecting two zigzag graphene
nanoribbons by using the non-equilirium Green's function approach combined with
the density functional theory calculation. The effects on the transport from
different spin configurations of the leads, different positions of the bridge
connection, and the width of the leads are investigated. 

It was found that the bridge position combined with the spin
freedom dominate the transport properties.  
A bridge connection in the middle will give an overall good coupling and
transparency because of the strong coupling between the bulk states in the leads
and those in the atomic chain, except for the energies around the Fermi energy
where the lead with anti-parallel spins creates a transport gap while the lead
with parallel spins give an finite density of states because of its band
crossing. The coupling between this finite density of states and the states in
the carbon chain gives rise to a large (for odd-numbered chains) or a small (for
even-numbered chains) equilibrium conductance for both the $\alpha$- and
$\beta$-spin components, inducing an even-odd oscillation in the equilibrium
conductance. On the other hand, a bridge at the edge leads to a transport
behavior associated with the edge states, showing only sharp pure $\alpha$-spin and
$\beta$-spin peaks beside the Fermi energy with a near zero equilibrium
conductance. 

Our calculation reveals that a functional device for on-chip interconnects can
be realized by a bridge connection in the near middle (M1-bridge), which gives a large
equilibrium conductance when the leads are in the parallel-spin
configuration. This spin configuration together with a bridge connected not at the
edge may also be used to realize spintronic device with a moderate
spin-polarization ratio. On the other hand, a functional spintronic device 
with a very high spin polarization ratio may be realized by a bridge connection at the edge with a small gate
voltage shifting the Fermi energy to the energy of the pure $\alpha$- or
$\beta$-spin peak. 

\begin{acknowledgments}
This work was supported by Shanghai Pujiang Program under Grant No. 10PJ1410000 and
by the National Natural Science Foundation of China under Grant No. 11174220
as well as by the MOST 973 Project 2011CB922204.
\end{acknowledgments}


\begin{thebibliography}{35}
\expandafter\ifx\csname natexlab\endcsname\relax\def\natexlab#1{#1}\fi
\expandafter\ifx\csname bibnamefont\endcsname\relax
  \def\bibnamefont#1{#1}\fi
\expandafter\ifx\csname bibfnamefont\endcsname\relax
  \def\bibfnamefont#1{#1}\fi
\expandafter\ifx\csname citenamefont\endcsname\relax
  \def\citenamefont#1{#1}\fi
\expandafter\ifx\csname url\endcsname\relax
  \def\url#1{\texttt{#1}}\fi
\expandafter\ifx\csname urlprefix\endcsname\relax\def\urlprefix{URL }\fi
\providecommand{\bibinfo}[2]{#2}
\providecommand{\eprint}[2][]{\url{#2}}

\bibitem[{\citenamefont{Avouris et~al.}(2007)\citenamefont{Avouris, Chen, and
  Perebeinos}}]{avouris2007carbon}
\bibinfo{author}{\bibfnamefont{P.}~\bibnamefont{Avouris}},
  \bibinfo{author}{\bibfnamefont{Z.}~\bibnamefont{Chen}}, \bibnamefont{and}
  \bibinfo{author}{\bibfnamefont{V.}~\bibnamefont{Perebeinos}},
  \bibinfo{journal}{Nat. Nanotechnol.} \textbf{\bibinfo{volume}{2}},
  \bibinfo{pages}{605} (\bibinfo{year}{2007}).

\bibitem[{\citenamefont{Nakada et~al.}(1996)\citenamefont{Nakada, Fujita,
  Dresselhaus, and Dresselhaus}}]{PhysRevB.54.17954}
\bibinfo{author}{\bibfnamefont{K.}~\bibnamefont{Nakada}},
  \bibinfo{author}{\bibfnamefont{M.}~\bibnamefont{Fujita}},
  \bibinfo{author}{\bibfnamefont{G.}~\bibnamefont{Dresselhaus}},
  \bibnamefont{and} \bibinfo{author}{\bibfnamefont{M.~S.}
  \bibnamefont{Dresselhaus}}, \bibinfo{journal}{Phys. Rev. B}
  \textbf{\bibinfo{volume}{54}}, \bibinfo{pages}{17954} (\bibinfo{year}{1996}).

\bibitem[{\citenamefont{Castro~Neto et~al.}(2009)\citenamefont{Castro~Neto,
  Guinea, Peres, Novoselov, and Geim}}]{RevModPhys.81.109}
\bibinfo{author}{\bibfnamefont{A.~H.} \bibnamefont{Castro~Neto}},
  \bibinfo{author}{\bibfnamefont{F.}~\bibnamefont{Guinea}},
  \bibinfo{author}{\bibfnamefont{N.~M.~R.} \bibnamefont{Peres}},
  \bibinfo{author}{\bibfnamefont{K.~S.} \bibnamefont{Novoselov}},
  \bibnamefont{and} \bibinfo{author}{\bibfnamefont{A.~K.} \bibnamefont{Geim}},
  \bibinfo{journal}{Rev. Mod. Phys.} \textbf{\bibinfo{volume}{81}},
  \bibinfo{pages}{109} (\bibinfo{year}{2009}).

\bibitem[{\citenamefont{Geim and Novoselov}(2007)}]{geim2007rise}
\bibinfo{author}{\bibfnamefont{A.}~\bibnamefont{Geim}} \bibnamefont{and}
  \bibinfo{author}{\bibfnamefont{K.}~\bibnamefont{Novoselov}},
  \bibinfo{journal}{Nat. Mater.} \textbf{\bibinfo{volume}{6}},
  \bibinfo{pages}{183} (\bibinfo{year}{2007}).

\bibitem[{\citenamefont{Han et~al.}(2007)\citenamefont{Han, \"Ozyilmaz, Zhang,
  and Kim}}]{PhysRevLett.98.206805}
\bibinfo{author}{\bibfnamefont{M.~Y.} \bibnamefont{Han}},
  \bibinfo{author}{\bibfnamefont{B.}~\bibnamefont{\"Ozyilmaz}},
  \bibinfo{author}{\bibfnamefont{Y.}~\bibnamefont{Zhang}}, \bibnamefont{and}
  \bibinfo{author}{\bibfnamefont{P.}~\bibnamefont{Kim}},
  \bibinfo{journal}{Phys. Rev. Lett.} \textbf{\bibinfo{volume}{98}},
  \bibinfo{pages}{206805} (\bibinfo{year}{2007}).

\bibitem[{\citenamefont{Son et~al.}(2006{\natexlab{a}})\citenamefont{Son,
  Cohen, and Louie}}]{PhysRevLett.97.216803}
\bibinfo{author}{\bibfnamefont{Y.-W.} \bibnamefont{Son}},
  \bibinfo{author}{\bibfnamefont{M.~L.} \bibnamefont{Cohen}}, \bibnamefont{and}
  \bibinfo{author}{\bibfnamefont{S.~G.} \bibnamefont{Louie}},
  \bibinfo{journal}{Phys. Rev. Lett.} \textbf{\bibinfo{volume}{97}},
  \bibinfo{pages}{216803} (\bibinfo{year}{2006}{\natexlab{a}}).

\bibitem[{\citenamefont{Peres et~al.}(2006)\citenamefont{Peres, Castro~Neto,
  and Guinea}}]{PhysRevB.73.195411}
\bibinfo{author}{\bibfnamefont{N.~M.~R.} \bibnamefont{Peres}},
  \bibinfo{author}{\bibfnamefont{A.~H.} \bibnamefont{Castro~Neto}},
  \bibnamefont{and} \bibinfo{author}{\bibfnamefont{F.}~\bibnamefont{Guinea}},
  \bibinfo{journal}{Phys. Rev. B} \textbf{\bibinfo{volume}{73}},
  \bibinfo{pages}{195411} (\bibinfo{year}{2006}).

\bibitem[{\citenamefont{Wang et~al.}(2008)\citenamefont{Wang, Ouyang, Li, Wang,
  Guo, and Dai}}]{PhysRevLett.100.206803}
\bibinfo{author}{\bibfnamefont{X.}~\bibnamefont{Wang}},
  \bibinfo{author}{\bibfnamefont{Y.}~\bibnamefont{Ouyang}},
  \bibinfo{author}{\bibfnamefont{X.}~\bibnamefont{Li}},
  \bibinfo{author}{\bibfnamefont{H.}~\bibnamefont{Wang}},
  \bibinfo{author}{\bibfnamefont{J.}~\bibnamefont{Guo}}, \bibnamefont{and}
  \bibinfo{author}{\bibfnamefont{H.}~\bibnamefont{Dai}},
  \bibinfo{journal}{Phys. Rev. Lett.} \textbf{\bibinfo{volume}{100}},
  \bibinfo{pages}{206803} (\bibinfo{year}{2008}).

\bibitem[{\citenamefont{Yan et~al.}(2007)\citenamefont{Yan, Huang, Yu, Zheng,
  Zang, Wu, Gu, Liu, and Duan}}]{doi:10.1021/nl070133j}
\bibinfo{author}{\bibfnamefont{Q.}~\bibnamefont{Yan}},
  \bibinfo{author}{\bibfnamefont{B.}~\bibnamefont{Huang}},
  \bibinfo{author}{\bibfnamefont{J.}~\bibnamefont{Yu}},
  \bibinfo{author}{\bibfnamefont{F.}~\bibnamefont{Zheng}},
  \bibinfo{author}{\bibfnamefont{J.}~\bibnamefont{Zang}},
  \bibinfo{author}{\bibfnamefont{J.}~\bibnamefont{Wu}},
  \bibinfo{author}{\bibfnamefont{B.-L.} \bibnamefont{Gu}},
  \bibinfo{author}{\bibfnamefont{F.}~\bibnamefont{Liu}}, \bibnamefont{and}
  \bibinfo{author}{\bibfnamefont{W.}~\bibnamefont{Duan}},
  \bibinfo{journal}{Nano Lett.} \textbf{\bibinfo{volume}{7}},
  \bibinfo{pages}{1469} (\bibinfo{year}{2007}).

\bibitem[{\citenamefont{Standley et~al.}(2008)\citenamefont{Standley, Bao,
  Zhang, Bruck, Lau, and Bockrath}}]{doi:10.1021/nl801774a}
\bibinfo{author}{\bibfnamefont{B.}~\bibnamefont{Standley}},
  \bibinfo{author}{\bibfnamefont{W.}~\bibnamefont{Bao}},
  \bibinfo{author}{\bibfnamefont{H.}~\bibnamefont{Zhang}},
  \bibinfo{author}{\bibfnamefont{J.}~\bibnamefont{Bruck}},
  \bibinfo{author}{\bibfnamefont{C.~N.} \bibnamefont{Lau}}, \bibnamefont{and}
  \bibinfo{author}{\bibfnamefont{M.}~\bibnamefont{Bockrath}},
  \bibinfo{journal}{Nano Lett.} \textbf{\bibinfo{volume}{8}},
  \bibinfo{pages}{3345} (\bibinfo{year}{2008}).

\bibitem[{\citenamefont{Son et~al.}(2006{\natexlab{b}})\citenamefont{Son,
  Cohen, and Louie}}]{son2006half}
\bibinfo{author}{\bibfnamefont{Y.}~\bibnamefont{Son}},
  \bibinfo{author}{\bibfnamefont{M.}~\bibnamefont{Cohen}}, \bibnamefont{and}
  \bibinfo{author}{\bibfnamefont{S.}~\bibnamefont{Louie}},
  \bibinfo{journal}{Nature} \textbf{\bibinfo{volume}{444}},
  \bibinfo{pages}{347} (\bibinfo{year}{2006}{\natexlab{b}}).

\bibitem[{\citenamefont{Meyer et~al.}(2008)\citenamefont{Meyer, Girit, Crommie,
  and Zettl}}]{meyer2008imaging}
\bibinfo{author}{\bibfnamefont{J.}~\bibnamefont{Meyer}},
  \bibinfo{author}{\bibfnamefont{C.}~\bibnamefont{Girit}},
  \bibinfo{author}{\bibfnamefont{M.}~\bibnamefont{Crommie}}, \bibnamefont{and}
  \bibinfo{author}{\bibfnamefont{A.}~\bibnamefont{Zettl}},
  \bibinfo{journal}{Nature} \textbf{\bibinfo{volume}{454}},
  \bibinfo{pages}{319} (\bibinfo{year}{2008}).

\bibitem[{\citenamefont{Chuvilin et~al.}(2009)\citenamefont{Chuvilin, Meyer,
  Algara-Siller, and Kaiser}}]{chuvilin2009graphene}
\bibinfo{author}{\bibfnamefont{A.}~\bibnamefont{Chuvilin}},
  \bibinfo{author}{\bibfnamefont{J.}~\bibnamefont{Meyer}},
  \bibinfo{author}{\bibfnamefont{G.}~\bibnamefont{Algara-Siller}},
  \bibnamefont{and} \bibinfo{author}{\bibfnamefont{U.}~\bibnamefont{Kaiser}},
  \bibinfo{journal}{New J. Phys.} \textbf{\bibinfo{volume}{11}},
  \bibinfo{pages}{083019} (\bibinfo{year}{2009}).

\bibitem[{\citenamefont{Jin et~al.}(2009)\citenamefont{Jin, Lan, Peng, Suenaga,
  and Iijima}}]{jin2009deriving}
\bibinfo{author}{\bibfnamefont{C.}~\bibnamefont{Jin}},
  \bibinfo{author}{\bibfnamefont{H.}~\bibnamefont{Lan}},
  \bibinfo{author}{\bibfnamefont{L.}~\bibnamefont{Peng}},
  \bibinfo{author}{\bibfnamefont{K.}~\bibnamefont{Suenaga}}, \bibnamefont{and}
  \bibinfo{author}{\bibfnamefont{S.}~\bibnamefont{Iijima}},
  \bibinfo{journal}{Phys. Rev. Lett.} \textbf{\bibinfo{volume}{102}},
  \bibinfo{pages}{205501} (\bibinfo{year}{2009}).

\bibitem[{\citenamefont{Basch et~al.}(2005)\citenamefont{Basch, Cohen, and
  Ratner}}]{Basch051668}
\bibinfo{author}{\bibfnamefont{H.}~\bibnamefont{Basch}},
  \bibinfo{author}{\bibfnamefont{R.}~\bibnamefont{Cohen}}, \bibnamefont{and}
  \bibinfo{author}{\bibfnamefont{M.}~\bibnamefont{Ratner}},
  \bibinfo{journal}{Nano Lett.} \textbf{\bibinfo{volume}{5}},
  \bibinfo{pages}{1668} (\bibinfo{year}{2005}).

\bibitem[{\citenamefont{Venkataraman et~al.}(2006)\citenamefont{Venkataraman,
  Klare, Tam, Nuckolls, Hybertsen, and Steigerwald}}]{Venkataraman06458}
\bibinfo{author}{\bibfnamefont{L.}~\bibnamefont{Venkataraman}},
  \bibinfo{author}{\bibfnamefont{J.}~\bibnamefont{Klare}},
  \bibinfo{author}{\bibfnamefont{I.}~\bibnamefont{Tam}},
  \bibinfo{author}{\bibfnamefont{C.}~\bibnamefont{Nuckolls}},
  \bibinfo{author}{\bibfnamefont{M.}~\bibnamefont{Hybertsen}},
  \bibnamefont{and}
  \bibinfo{author}{\bibfnamefont{M.}~\bibnamefont{Steigerwald}},
  \bibinfo{journal}{Nano Lett.} \textbf{\bibinfo{volume}{6}},
  \bibinfo{pages}{458} (\bibinfo{year}{2006}).

\bibitem[{\citenamefont{Ke et~al.}(2005)\citenamefont{Ke, Baranger, and
  Yang}}]{Ke05074704}
\bibinfo{author}{\bibfnamefont{S.-H.} \bibnamefont{Ke}},
  \bibinfo{author}{\bibfnamefont{H.}~\bibnamefont{Baranger}}, \bibnamefont{and}
  \bibinfo{author}{\bibfnamefont{W.}~\bibnamefont{Yang}}, \bibinfo{journal}{J.
  Chem. Phys.} \textbf{\bibinfo{volume}{122}}, \bibinfo{pages}{074704}
  (\bibinfo{year}{2005}).

\bibitem[{\citenamefont{Tongay et~al.}(2004)\citenamefont{Tongay, Senger, Dag,
  and Ciraci}}]{tongay2004ab}
\bibinfo{author}{\bibfnamefont{S.}~\bibnamefont{Tongay}},
  \bibinfo{author}{\bibfnamefont{R.}~\bibnamefont{Senger}},
  \bibinfo{author}{\bibfnamefont{S.}~\bibnamefont{Dag}}, \bibnamefont{and}
  \bibinfo{author}{\bibfnamefont{S.}~\bibnamefont{Ciraci}},
  \bibinfo{journal}{Phys. Rev. Lett.} \textbf{\bibinfo{volume}{93}},
  \bibinfo{pages}{136404} (\bibinfo{year}{2004}).

\bibitem[{\citenamefont{Lang and Avouris}(2000)}]{lang2000carbon}
\bibinfo{author}{\bibfnamefont{N.}~\bibnamefont{Lang}} \bibnamefont{and}
  \bibinfo{author}{\bibfnamefont{P.}~\bibnamefont{Avouris}},
  \bibinfo{journal}{Phys. Rev. Lett.} \textbf{\bibinfo{volume}{84}},
  \bibinfo{pages}{358} (\bibinfo{year}{2000}).

\bibitem[{\citenamefont{Zhou et~al.}(2008)\citenamefont{Zhou, Zheng, Xu, and
  Zeng}}]{zhou2008first}
\bibinfo{author}{\bibfnamefont{Y.}~\bibnamefont{Zhou}},
  \bibinfo{author}{\bibfnamefont{X.}~\bibnamefont{Zheng}},
  \bibinfo{author}{\bibfnamefont{Y.}~\bibnamefont{Xu}}, \bibnamefont{and}
  \bibinfo{author}{\bibfnamefont{Z.}~\bibnamefont{Zeng}}, \bibinfo{journal}{J.
  Phys.: Condens. Matter.} \textbf{\bibinfo{volume}{20}},
  \bibinfo{pages}{045225} (\bibinfo{year}{2008}).

\bibitem[{\citenamefont{Brandbyge et~al.}(2002)\citenamefont{Brandbyge, Mozos,
  Ordejon, Taylor, and Stokbro}}]{brandbyge2002density}
\bibinfo{author}{\bibfnamefont{M.}~\bibnamefont{Brandbyge}},
  \bibinfo{author}{\bibfnamefont{J.}~\bibnamefont{Mozos}},
  \bibinfo{author}{\bibfnamefont{P.}~\bibnamefont{Ordejon}},
  \bibinfo{author}{\bibfnamefont{J.}~\bibnamefont{Taylor}}, \bibnamefont{and}
  \bibinfo{author}{\bibfnamefont{K.}~\bibnamefont{Stokbro}},
  \bibinfo{journal}{Phys. Rev. B} \textbf{\bibinfo{volume}{65}},
  \bibinfo{pages}{165401} (\bibinfo{year}{2002}).

\bibitem[{\citenamefont{Larade et~al.}(2001)\citenamefont{Larade, Taylor,
  Mehrez, and Guo}}]{larade2001conductance}
\bibinfo{author}{\bibfnamefont{B.}~\bibnamefont{Larade}},
  \bibinfo{author}{\bibfnamefont{J.}~\bibnamefont{Taylor}},
  \bibinfo{author}{\bibfnamefont{H.}~\bibnamefont{Mehrez}}, \bibnamefont{and}
  \bibinfo{author}{\bibfnamefont{H.}~\bibnamefont{Guo}},
  \bibinfo{journal}{Phys. Rev. B} \textbf{\bibinfo{volume}{64}},
  \bibinfo{pages}{075420} (\bibinfo{year}{2001}).

\bibitem[{\citenamefont{Wei et~al.}(2004)\citenamefont{Wei, Xu, Wang, and
  Guo}}]{wei2004spin}
\bibinfo{author}{\bibfnamefont{Y.}~\bibnamefont{Wei}},
  \bibinfo{author}{\bibfnamefont{Y.}~\bibnamefont{Xu}},
  \bibinfo{author}{\bibfnamefont{J.}~\bibnamefont{Wang}}, \bibnamefont{and}
  \bibinfo{author}{\bibfnamefont{H.}~\bibnamefont{Guo}},
  \bibinfo{journal}{Phys. Rev. B} \textbf{\bibinfo{volume}{70}},
  \bibinfo{pages}{193406} (\bibinfo{year}{2004}).

\bibitem[{\citenamefont{Khoo et~al.}(2008)\citenamefont{Khoo, Neaton, Son,
  Cohen, and Louie}}]{khoo2008negative}
\bibinfo{author}{\bibfnamefont{K.}~\bibnamefont{Khoo}},
  \bibinfo{author}{\bibfnamefont{J.}~\bibnamefont{Neaton}},
  \bibinfo{author}{\bibfnamefont{Y.}~\bibnamefont{Son}},
  \bibinfo{author}{\bibfnamefont{M.}~\bibnamefont{Cohen}}, \bibnamefont{and}
  \bibinfo{author}{\bibfnamefont{S.}~\bibnamefont{Louie}},
  \bibinfo{journal}{Nano Lett.} \textbf{\bibinfo{volume}{8}},
  \bibinfo{pages}{2900} (\bibinfo{year}{2008}).

\bibitem[{\citenamefont{Cheraghchi and
  Esfarjani}(2008)}]{cheraghchi2008negative}
\bibinfo{author}{\bibfnamefont{H.}~\bibnamefont{Cheraghchi}} \bibnamefont{and}
  \bibinfo{author}{\bibfnamefont{K.}~\bibnamefont{Esfarjani}},
  \bibinfo{journal}{Phys. Rev. B} \textbf{\bibinfo{volume}{78}},
  \bibinfo{pages}{085123} (\bibinfo{year}{2008}).

\bibitem[{\citenamefont{F{\"u}rst et~al.}(2010)\citenamefont{F{\"u}rst,
  Brandbyge, and Jauho}}]{furst2010atomic}
\bibinfo{author}{\bibfnamefont{J.}~\bibnamefont{F{\"u}rst}},
  \bibinfo{author}{\bibfnamefont{M.}~\bibnamefont{Brandbyge}},
  \bibnamefont{and} \bibinfo{author}{\bibfnamefont{A.}~\bibnamefont{Jauho}},
  \bibinfo{journal}{Europhys. Lett.} \textbf{\bibinfo{volume}{91}},
  \bibinfo{pages}{37002} (\bibinfo{year}{2010}).

\bibitem[{\citenamefont{Ke et~al.}(2007)\citenamefont{Ke, Baranger, and
  Yang}}]{Ke07146802}
\bibinfo{author}{\bibfnamefont{S.-H.} \bibnamefont{Ke}},
  \bibinfo{author}{\bibfnamefont{H.}~\bibnamefont{Baranger}}, \bibnamefont{and}
  \bibinfo{author}{\bibfnamefont{W.}~\bibnamefont{Yang}},
  \bibinfo{journal}{Phy. Rev. Lett.} \textbf{\bibinfo{volume}{99}},
  \bibinfo{pages}{146802} (\bibinfo{year}{2007}).

\bibitem[{\citenamefont{Shen et~al.}(2010)\citenamefont{Shen, Zeng, Yang,
  Zhang, Wang, and Feng}}]{shen2010electron}
\bibinfo{author}{\bibfnamefont{L.}~\bibnamefont{Shen}},
  \bibinfo{author}{\bibfnamefont{M.}~\bibnamefont{Zeng}},
  \bibinfo{author}{\bibfnamefont{S.}~\bibnamefont{Yang}},
  \bibinfo{author}{\bibfnamefont{C.}~\bibnamefont{Zhang}},
  \bibinfo{author}{\bibfnamefont{X.}~\bibnamefont{Wang}}, \bibnamefont{and}
  \bibinfo{author}{\bibfnamefont{Y.}~\bibnamefont{Feng}}, \bibinfo{journal}{J.
  Am. Chem. Soc.}  (\bibinfo{year}{2010}).

\bibitem[{\citenamefont{Zanolli et~al.}(2010)\citenamefont{Zanolli, Onida, and
  Charlier}}]{zanolli2010quantum}
\bibinfo{author}{\bibfnamefont{Z.}~\bibnamefont{Zanolli}},
  \bibinfo{author}{\bibfnamefont{G.}~\bibnamefont{Onida}}, \bibnamefont{and}
  \bibinfo{author}{\bibfnamefont{J.}~\bibnamefont{Charlier}},
  \bibinfo{journal}{ACS nano} \textbf{\bibinfo{volume}{4}},
  \bibinfo{pages}{5174} (\bibinfo{year}{2010}).

\bibitem[{\citenamefont{Datta}(1995)}]{datta95}
\bibinfo{author}{\bibfnamefont{S.}~\bibnamefont{Datta}},
  \emph{\bibinfo{title}{Electronic Transport in Mesoscopic Systems}}
  (\bibinfo{publisher}{Cambridge University Press},
  \bibinfo{address}{Cambridge, England}, \bibinfo{year}{1995}).

\bibitem[{\citenamefont{Ke et~al.}(2004)\citenamefont{Ke, Baranger, and
  Yang}}]{ke2004electron}
\bibinfo{author}{\bibfnamefont{S.}~\bibnamefont{Ke}},
  \bibinfo{author}{\bibfnamefont{H.}~\bibnamefont{Baranger}}, \bibnamefont{and}
  \bibinfo{author}{\bibfnamefont{W.}~\bibnamefont{Yang}},
  \bibinfo{journal}{Phys. Rev. B} \textbf{\bibinfo{volume}{70}},
  \bibinfo{pages}{085410} (\bibinfo{year}{2004}).

\bibitem[{\citenamefont{Soler et~al.}(2002)\citenamefont{Soler, Artacho, Gale,
  Garc{\'\i}a, Junquera, Ordej{\'o}n, and
  S{\'a}nchez-Portal}}]{soler2002siesta}
\bibinfo{author}{\bibfnamefont{J.}~\bibnamefont{Soler}},
  \bibinfo{author}{\bibfnamefont{E.}~\bibnamefont{Artacho}},
  \bibinfo{author}{\bibfnamefont{J.}~\bibnamefont{Gale}},
  \bibinfo{author}{\bibfnamefont{A.}~\bibnamefont{Garc{\'\i}a}},
  \bibinfo{author}{\bibfnamefont{J.}~\bibnamefont{Junquera}},
  \bibinfo{author}{\bibfnamefont{P.}~\bibnamefont{Ordej{\'o}n}},
  \bibnamefont{and}
  \bibinfo{author}{\bibfnamefont{D.}~\bibnamefont{S{\'a}nchez-Portal}},
  \bibinfo{journal}{J. Phys.: Condens. Matter.} \textbf{\bibinfo{volume}{14}},
  \bibinfo{pages}{2745} (\bibinfo{year}{2002}).

\bibitem[{\citenamefont{Ceperley and Alder}(1980)}]{PhysRevLett.45.566}
\bibinfo{author}{\bibfnamefont{D.~M.} \bibnamefont{Ceperley}} \bibnamefont{and}
  \bibinfo{author}{\bibfnamefont{B.~J.} \bibnamefont{Alder}},
  \bibinfo{journal}{Phys. Rev. Lett.} \textbf{\bibinfo{volume}{45}},
  \bibinfo{pages}{566} (\bibinfo{year}{1980}).

\bibitem[{\citenamefont{Troullier and Martins}(1991)}]{Troullier911993}
\bibinfo{author}{\bibfnamefont{N.}~\bibnamefont{Troullier}} \bibnamefont{and}
  \bibinfo{author}{\bibfnamefont{J.~L.} \bibnamefont{Martins}},
  \bibinfo{journal}{Phy. Rev. B} \textbf{\bibinfo{volume}{43}},
  \bibinfo{pages}{1993} (\bibinfo{year}{1991}).

\bibitem[{spi()}]{spin-polarization}
\bibinfo{note}{The converged spin-polarized electronic states in the carbon
  chain are sensitive to the choice for the initial spins of the atoms at the
  edges of the leads. Therefore these initial spins must be chosen carefully
  otherwise it may be converged to the spin-unpolarized state. The total energy
  of the system shows that the spin-polarized state is the ground state.}

\end{thebibliography}

\end{document}